\documentclass[12pt, authoryear]{article}
\usepackage{amsfonts, graphicx,color,multirow, amsthm,amsmath,natbib}
\usepackage{setspace,url}
\usepackage{multirow}

\def\boxit#1{\vbox{\hrule\hbox{\vrule\kern6pt\vbox{\kern6pt#1\kern6pt}\kern6pt\vrule}\hrule}}

\begin{document}
\newtheorem{theorem}{Theorem}
\newtheorem{corollary}{Corollary}
\newtheorem{case}{Case}
\newtheorem{definition}{Definition}
\newtheorem{example}{Example}
\newtheorem{lemma}{Lemma}
\newtheorem{proposition}{Proposition}
\newtheorem{remark}{Remark}
\newtheorem{assumption}{Assumption}

\newcommand{\ol}[1]{\overline{#1}}

\newcommand{\bb}{\mbox{\bf b}}
\newcommand{\bff}{\mbox{\bf f}}
\newcommand{\bx}{\mbox{\bf x}}
\newcommand{\by}{\mbox{\bf y}}
\newcommand{\bA}{\mbox{\bf A}}
\newcommand{\ba}{\mbox{\bf a}}
\newcommand{\bw}{\mbox{\bf w}}
\newcommand{\bu}{\mbox{\bf u}}
\newcommand{\bB}{\mbox{\bf B}}
\newcommand{\bC}{\mbox{\bf C}}
\newcommand{\bD}{\mbox{\bf D}}
\newcommand{\bE}{\mbox{\bf E}}
\newcommand{\bF}{\mbox{\bf F}}
\newcommand{\bG}{\mbox{\bf G}}
\newcommand{\bH}{\mbox{\bf H}}
\newcommand{\bI}{\mbox{\bf I}}
\newcommand{\bK}{\mbox{\bf K}}
\newcommand{\bL}{\mbox{\bf L}}
\newcommand{\bQ}{\mbox{\bf Q}}
\newcommand{\bR}{\mbox{\bf R}}
\newcommand{\bS}{\mbox{\bf S}}
\newcommand{\bU}{\mbox{\bf U}}
\newcommand{\bX}{\mbox{\bf X}}
\newcommand{\bW}{\mbox{\bf W}}
\newcommand{\bY}{\mbox{\bf Y}}
\newcommand{\bZ}{\mbox{\bf Z}}
\newcommand{\bone}{\mbox{\bf 1}}
\newcommand{\bsone}{\mbox{\bf 1}}
\newcommand{\bzero}{\mbox{\bf 0}}
\newcommand{\bveps}{\mbox{\boldmath $\varepsilon$}}
\newcommand{\bet}{\mbox{\boldmath $\eta$}}
\newcommand{\bxi}{\mbox{\boldmath $b$}}
\newcommand{\beps}{\mbox{\boldmath $\varepsilon$}}
\newcommand{\bmu}{\mbox{\boldmath $\mu$}}
\newcommand{\bgamma}{\mbox{\boldmath $\gamma$}}
\newcommand{\bbeta}{\mbox{\boldmath $\beta$}}
\newcommand{\bGamma}{\mbox{\boldmath $\Gamma$}}
\newcommand{\mv}{\mbox{V}}
\newcommand{\bSigma}{\mbox{\boldmath $\Sigma$}}
\newcommand{\bOmega}{\mbox{\boldmath $\Omega$}}
\newcommand{\FDR}{\mbox{FDR}}
\newcommand{\FDP}{\mbox{FDP}}
\newcommand{\FNR}{\mbox{FNR}}
\newcommand{\wFDP}{\mbox{$\widehat{FDP}$}}
\newcommand{\wFDR}{\mbox{$\widehat{FDR}$}}
\newcommand{\LS}{\mbox{\scriptsize LS}}
\newcommand{\Ll}{\mbox{\scriptsize $L_1$}}
\newcommand{\Var}{\mbox{Var}}
\newcommand{\Cov}{\mbox{Cov}}
\newcommand{\SNP}{\mbox{SNP}}
\newcommand{\SD}{\mbox{SD}}

\newcommand{\hB}{\widehat \bB}
\newcommand{\hb}{\widehat \bb}
\newcommand{\hw}{\widehat \bw}
\newcommand{\hE}{\widehat \bE}
\newcommand{\hvar}{\widehat \var}
\newcommand{\hcov}{\widehat \cov}
\newcommand{\hbveps}{\widehat\bveps}
\newcommand{\hSig}{\widehat\Sig}
\newcommand{\hsig}{\widehat\sigma}
\newcommand{\hmu}{\widehat\bmu}
\newcommand{\hxi}{\widehat\bxi}
\newcommand{\heq}{\ \widehat=\ }
\newcommand{\sam}{_{\text{sam}}}
\newcommand{\cov}{\mathrm{cov}}
\newcommand{\Sig}{\mathbf{\Sigma}}
\newcommand{\veps}{\varepsilon}
\newcommand{\tr}{\mathrm{tr}}
\newcommand{\diag}{\mathrm{diag}}
\newcommand{\argmin}{\mathrm{argmin}}
\newcommand{\vecc}{\mathrm{vec}}
\newcommand{\var}{\mathrm{var}}
\def\cov{\mbox{cov}}
\newcommand{\trace}{\mathrm{tr}}

\def\t{^T}
\def\toD{\overset{\mathrm{D}}{\longrightarrow}}
\def\toP{\overset{\mathrm{P}}{\longrightarrow}}
\def\toas{\overset{\mathrm{a.s.}}{\longrightarrow}}
\def\deq{\overset{\mathrm{(d)}}{=\hspace{-0.02 in}=}}

\def\blackbox{\vrule height4pt width3pt depth0pt}
\def\indep{\perp \!\!\!\!\perp}
\def\bB{\mbox{\bf B}}

\title{\bf Estimating False Discovery Proportion Under Arbitrary Covariance Dependence\thanks{Jianqing Fan is Frederick L. Moore'18 professor, Department of Operations Research \& Financial Engineering, Princeton University, Princeton, NJ 08544, USA and honorary professor, School of Statistics and Management, Shanghai University of Finance and Economics, Shanghai, China (Email: jqfan@princeton.edu).
Xu Han is assistant professor, Department of Statistics, University of Florida, Florida, FL 32606 (Email: xhan@princeton.edu).  Weijie Gu is graduate student, Department of Operations Research \& Financial Engineering, Princeton University, Princeton, NJ 08544 (Email: wgu@princeton.edu).  The paper was completed while Xu Han was a postdoctoral fellow at Princeton University.  This research was partly supported by NSF Grants DMS-0704337 and DMS-0714554 and NIH Grant R01-GM072611. The authors are grateful to the editor, associate editor and referees for helpful comments.}
}

\author{Jianqing Fan, Xu Han and Weijie Gu}
\maketitle

\begin{abstract}
\begin{singlespace}
Multiple hypothesis testing is a fundamental problem in high dimensional inference, with wide applications in many scientific fields. In genome-wide association studies, tens of thousands of tests are performed simultaneously to find if any SNPs are associated with some traits and those tests are correlated. When test statistics are correlated, false discovery control becomes very challenging under arbitrary dependence. In the current paper, we propose a novel method based on principal factor approximation, which successfully subtracts the common dependence and weakens significantly the correlation structure, to deal with an arbitrary dependence structure. We derive an approximate expression for false discovery proportion (FDP) in large scale multiple testing when a common threshold is used and provide a consistent estimate of realized FDP. This result has important applications in controlling FDR and FDP. Our estimate of realized FDP compares favorably with Efron (2007)'s approach, as demonstrated in the simulated examples. Our approach is further illustrated by some real data applications. We also propose a dependence-adjusted procedure, which is more powerful than the fixed threshold procedure.
\end{singlespace}
\end{abstract}

\section{Introduction}
Multiple hypothesis testing is a fundamental problem in the modern research for high dimensional inference, with wide applications in scientific fields, such as biology, medicine, genetics, neuroscience, economics and finance. For example, in genome-wide association studies, massive amount of genomic data (e.g. SNPs, eQTLs) are collected and tens of thousands of hypotheses are tested simultaneously to find if any of these genomic data are associated with some observable traits (e.g. blood pressure, weight, some disease); in finance, thousands of tests are performed to see which fund managers have winning ability (Barras, Scaillet \& Wermers 2010).

False Discovery Rate (FDR) has been introduced in the celebrated paper by Benjamini \& Hochberg (1995) for large scale multiple testing. By definition, FDR is the expected proportion of falsely rejected null hypotheses among all of the rejected null hypotheses. The classification of tested hypotheses can be summarized in Table 1:
\begin{table}[h!!]
\begin{center}\caption{Classification of tested hypotheses}\label{ar}
  \begin{tabular}{cccc}
         \hline\hline
                       &Number    &Number    &\\
         Number of     &not rejected  &rejected  &Total\\
         \hline
         True Null     &$U$         &$V$      &$p_0$\\
         False Null    &$T$         &$S$      &$p_1$\\
         \hline
                       &$p-R$       &$R$      &$p$\\
         \hline
   \end{tabular}
\end{center}
\end{table}

Various testing procedures have been developed for controlling FDR, among which there are two major approaches. One is to compare the $P$-values with a data-driven threshold as in Benjamini \& Hochberg (1995).  Specifically, let
$p_{(1)} \leq p_{(2)} \leq \cdots \leq p_{(p)}$ be the ordered observed $P$-values of $p$ hypotheses. Define $k = \text{max}\Big\{i: p_{(i)} \leq i\alpha/p \Big\}$ and reject $H_{(1)}^0, \cdots, H_{(k)}^0$, where $\alpha$ is a specified control rate. If no such $i$ exists, reject no hypothesis. The other related approach is to fix a threshold value $t$, estimate the FDR, and choose $t$ so that the estimated FDR is no larger than $\alpha$ (Storey 2002).  Finding such a common threshold is based on a conservative estimate of FDR. Specifically, let $\mathrm{\widehat{FDR}}(t) = \widehat{p}_0t/(R(t)\vee 1)$, where $R(t)$ is the number of total discoveries with the threshold $t$ and $\widehat{p}_0$ is an estimate of $p_0$. Then solve $t$ such that $\mathrm{\widehat{FDR}}(t)\leq\alpha$.  The equivalence between the two methods has been theoretically studied by Storey, Taylor \& Siegmund (2004) and Ferreira \& Zwinderman (2006).

Both procedures have been shown to control the FDR for independent test statistics. However, in practice, test statistics are usually correlated. Although Benjamini \& Yekutieli (2001) and Clarke \& Hall (2009) argued that when the null distribution of test statistics satisfies some conditions, dependence case in the multiple testing is asymptotically the same as independence case, multiple testing under general dependence structures is still a very challenging and important open problem. Efron (2007) pioneered the work in the field and noted that correlation must be accounted for in deciding which null hypotheses are significant because the accuracy of false discovery rate techniques will be compromised in high correlation situations. There are several papers that show the Benjamini-Hochberg procedure or Storey's procedure can control FDR under some special dependence structures, e.g. Positive Regression Dependence on Subsets (Benjamini \& Yekutieli 2001) and weak dependence (Storey, Taylor \& Siegmund 2004). Sarkar (2002) also shows that FDR can be controlled by a generalized stepwise multiple testing procedure under positive regression dependence on subsets. However, even if the procedures are valid under these special dependence structures, they will still suffer from efficiency loss without considering the actual dependence information. In other words, there are universal upper bounds for a given class of covariance matrices.

In the current paper, we will develop a procedure for high dimensional multiple testing which can deal with any arbitrary dependence structure and fully incorporate the covariance information. This is in contrast with previous literatures which consider multiple testing under special dependence structures, e.g. Sun \& Cai (2009) developed a multiple testing procedure where parameters underlying test statistics follow a hidden Markov model, and Leek \& Storey (2008) and Friguet, Kloareg \& Causeur (2009) studied multiple testing under the factor models. More specifically, consider the test statistics
\begin{equation*}
(Z_1,\cdots,Z_p)^T\sim N((\mu_1,\cdots,\mu_p)^T,\bSigma),
\end{equation*}
where $\bSigma$ is known and $p$ is large. We would like to simultaneously test $H_{0i}: \mu_i=0$ vs $H_{1i}: \mu_i\neq0$ for $i=1,\cdots,p$. Note that $\bSigma$ can be any non-negative definite matrix. Our procedure is called Principal Factor Approximation (PFA). The basic idea is to first take out the principal factors that derive the strong dependence among observed data $Z_1,\cdots,Z_p$ and to account for such dependence in calculation of false discovery proportion (FDP). This is accomplished by the spectral decomposition of $\bSigma$ and taking out the largest common factors so that the remaining dependence is weak. We then derive the asymptotic expression of the FDP, defined as $V/R$, that accounts for the strong dependence. The realized but unobserved principal factors that derive the strong dependence are then consistently estimated. The estimate of the realized FDP is obtained by substituting the consistent estimate of the unobserved principal factors.

We are especially interested in estimating FDP under the high dimensional sparse problem, that is, $p$ is very large, but the number of $\mu_i\neq0$ is very small. In section 2, we will explain the situation under which $\bSigma$ is known. Sections 3 and 4 present the theoretical results and the proposed procedures. In section 5, the performance of our procedures is critically evaluated by various simulation studies. Section 6 is about the real data analysis. All the proofs are relegated to the Appendix and the Supplemental Material.

\section{Motivation of the Study}
In genome-wide association studies, consider $p$ SNP genotype data for $n$ individual samples, and further suppose that a response of interest (i.e. gene expression level or a measure of phenotype such as blood pressure or weight) is recorded for each sample. The SNP data are conventionally stored in an $n\times p$ matrix $\bX=(x_{ij})$, with rows corresponding to individual samples and columns corresponding to individual SNPs . The total number $n$ of samples is in the order of hundreds, and the number $p$ of SNPs is in the order of tens of thousands.

Let $X_j$ and $Y$ denote, respectively, the random variables that correspond to the $j$th SNP coding and the outcome. The biological question of the association between genotype and phenotype can be restated as a problem in multiple hypothesis testing, {\em i.e.}, the simultaneous tests for each SNP $j$ of the null hypothesis $H_j$ of no association between the SNP $X_j$ and $Y$. Let $\{X_j^i\}_{i=1}^n$ be the sample data of $X_j$, $\{Y^i\}_{i=1}^n$ be the independent sample random variables of $Y$. Consider the marginal linear regression between $\{Y^i\}_{i=1}^n$ and $\{X_j^i\}_{i=1}^n$:
\begin{equation}\label{gwj1}
(\alpha_j,\beta_j)=\argmin_{a_j,b_j}\frac{1}{n}\sum_{i=1}^nE(Y^i-a_j-b_jX_j^i)^2, \ \ \ j=1,\cdots,p.
\end{equation}
We wish to simultaneously test the hypotheses
\begin{equation}
H_{0j}:\quad\beta_j=0\quad\text{vs}\quad H_{1j}:\quad\beta_j\neq0, \quad\quad j=1,\cdots,p
\end{equation}
to see which SNPs are correlated with the phenotype.

Recently statisticians have increasing interests in the high dimensional sparse problem: although the number of hypotheses to be tested is large, the number of false nulls ($\beta_j\neq0$) is very small. For example, among 2000 SNPs, there are maybe only 10 SNPs which contribute to the variation in phenotypes or certain gene expression level. Our purpose is to find these 10 SNPs by multiple testing with some statistical accuracy.

Because of the sample correlations among $\{X_j^i\}_{i=1,j=1}^{i=n,j=p}$, the least-squares estimators $\{\widehat{\beta}_j\}_{j=1}^p$ for $\{\beta_j\}_{j=1}^p$ in (1) are also correlated. The following result describes the joint distribution of $\{\widehat{\beta}_j\}_{j=1}^p$. The proof is straightforward.

{\proposition Let $\widehat{\beta}_j$ be the least-squares estimator for $\beta_j$ in (1) based on $n$ data points, $s_{kl}$ be the sample correlation between $X_k$ and $X_l$. Assume that the conditional distribution of $Y^i$ given $X_1^i,\cdots,X_p^i$ is $N(\mu(X_1^i,\cdots,X_p^i),\sigma^2)$. Then, conditioning on $\{X_j^i\}_{i=1,j=1}^{i=n,j=p}$, the joint distribution of $\{\widehat{\beta}_j\}_{j=1}^p$ is
$(\widehat{\beta}_1,\cdots,\widehat{\beta}_p)^T\sim N((\beta_1,\cdots,\beta_p)^T,\bSigma^*)$, where the $(k,l)$th element in $\bSigma^*$ is $\displaystyle\bSigma_{kl}^*=\sigma^2s_{kl}/(ns_{kk}s_{ll})$.}

For ease of notation, let $Z_1,\cdots,Z_p$ be the standardized random variables of $\widehat{\beta}_1,\cdots,\widehat{\beta}_p$, that is,
\begin{equation} \label{b2}
Z_i=\frac{\widehat{\beta}_i}{\SD(\widehat{\beta}_i)}=\frac{\widehat{\beta}_i}{\sigma/(\sqrt{n}s_{ii})}, \quad\quad i=1,\cdots,p.
\end{equation}
In the above, we implicitly assume that $\sigma$ is known and the above standardized random variables are z-test statistics.  The estimate of residual variance $\sigma^2$ will be discussed in Section 6 via refitted cross-validation (Fan, Guo \& Hao, 2011).  Then, conditioning on $\{X_j^i\}$,
\begin{equation}\label{c1}
(Z_1,\cdots,Z_p)^T\sim N((\mu_1,\cdots,\mu_p)^T,\bSigma),
\end{equation}
where $\mu_i=\sqrt{n}\beta_is_{ii}/\sigma$ and covariance matrix $\bSigma$ has the $(k,l)$th element as $s_{kl}$. Simultaneously testing (2) based on $(\widehat{\beta}_1,\cdots,\widehat{\beta}_p)^T$ is thus equivalent to testing

\begin{equation}\label{c2}
H_{0j}:\quad\mu_j=0\quad\text{vs}\quad H_{1j}:\quad \mu_j\neq0, \quad\quad j=1,\cdots,p
\end{equation}
based on $(Z_1,\cdots,Z_p)^T$.

In (4), $\bSigma$ is the population covariance matrix of $(Z_1,\cdots,Z_p)^T$, and is known based on the sample data $\{X_j^i\}$. The covariance matrix $\bSigma$ can have arbitrary dependence structure. We would like to clarify that $\bSigma$ is known and there is no estimation of the covariance matrix of $X_1,\cdots,X_p$ in this set up.

\section{Estimating False Discovery Proportion}
From now on assume that among all the $p$ null hypotheses, $p_0$ of them are true and $p_1$ hypotheses ($p_1 = p - p_0$) are
false, and $p_1$ is supposed to be very small compared to $p$. For ease of presentation, we let $p$ be the sole asymptotic parameter, and assume $p_0\rightarrow\infty$ when $p\rightarrow\infty$. For a fixed rejection threshold $t$, we will reject those $P$-values no greater than $t$ and regard them as statistically significance. Because of its powerful applicability, this procedure has been widely adopted; see, e.g., Storey (2002), Efron (2007, 2010), among others. Our goal is to estimate the realized FDP for a given $t$ in multiple testing problem (\ref{c2}) based on the observations (\ref{c1}) under arbitrary dependence structure of $\bSigma$.  Our methods and results have direct implications on the situation in which $\bSigma$ is unknown, the setting studied by Efron (2007, 2010) and Friguet et al (2009).  In the latter case, $\bSigma$ needs to be estimated with certain accuracy.

\subsection{Approximation of FDP}
Define the following empirical processes:
\begin{eqnarray*}
V(t) & = & \#\{true \ null \ P_i: P_i \leq t\}, \nonumber\\
S(t) & = & \#\{false \ null \ P_i: P_i \leq t\} \quad \text{and} \nonumber\\
R(t) & = & \#\{P_i: P_i \leq t\}, \nonumber
\end{eqnarray*}
where $t\in[0,1]$. Then, $V(t)$, $S(t)$ and $R(t)$ are the number of false discoveries, the number of true discoveries, and the number of total discoveries, respectively. Obviously, $R(t) = V(t) + S(t)$, and $V(t)$, $S(t)$ and $R(t)$ are all random variables, depending on the test statistics $(Z_1,\cdots,Z_p)^T$. Moreover, $R(t)$ is observed in an experiment, but $V(t)$ and $S(t)$ are both unobserved.

By definition, $\FDP(t)=V(t)/R(t)$ and $\FDR(t) = E\Big[V(t)/R(t)\Big]$. One of interests is to control FDR$(t)$ at a predetermined rate $\alpha$, say $15\%$. There are also substantial research interests in the statistical behavior of the number of false discoveries $V(t)$ and the false discovery proportion $\FDP(t)$, which are unknown but realized given an experiment.  One may even argue that controlling FDP is more relevant, since it is directly related to the current experiment.

We now approximate $V(t)/R(t)$ for the high dimensional sparse case $p_1\ll p$. Suppose $(Z_1,\cdots,Z_p)^T\sim N((\mu_1,\cdots,\mu_p)^T,\bSigma)$ in which $\bSigma$ is a correlation matrix. We need the following definition for weakly dependent normal random variables; other definitions can be found in Farcomeni (2007).

\begin{definition}
Suppose $(K_1,\cdots,K_p)^T\sim N((\theta_1,\cdots,\theta_p)^T,\bA)$. Then $K_1,\cdots,K_p$ are called weakly dependent normal variables if
\begin{equation}
\lim_{p\rightarrow\infty}p^{-2}\sum_{i,j}|a_{ij}|=0,
\end{equation}
where $a_{ij}$ denote the $(i,j)$th element of the covariance matrix $\bA$.
\end{definition}

Our procedure is called principal factor approximation (PFA). The basic idea is to decompose any dependent normal random vector as a factor model with weakly dependent normal random errors. The details are shown as follows.
Firstly apply the spectral decomposition to the covariance matrix $\bSigma$. Suppose the eigenvalues of $\bSigma$ are $\lambda_1,\cdots,\lambda_{p}$, which have been arranged in decreasing order. If the corresponding orthonormal eigenvectors are denoted as $\bgamma_1,\cdots,\bgamma_p$, then
\begin{equation}
\bSigma=\sum_{i=1}^p\lambda_i\bgamma_i\bgamma_i^T.
\end{equation}
If we further denote $\bA = \sum_{i=k+1}^p\lambda_i\bgamma_i\bgamma_i^T$  for an integer $k$, then
\begin{equation}
\|\bA\|_F^2 =\lambda_{k+1}^2+\cdots+\lambda_p^2,
\end{equation}
where $\|\cdot\|_F$ is the Frobenius norm. Let $\bL=(\sqrt{\lambda_1}\bgamma_1,\cdots,\sqrt{\lambda_k}\bgamma_k)$, which is a $p\times k$ matrix. Then the covariance matrix $\bSigma$ can be expressed as
\begin{equation} \label{b19}
\bSigma=\bL\bL^T+\bA,
\end{equation}
and $Z_1,\cdots,Z_p$ can be written as
\begin{equation} \label{b20}
Z_i=\mu_i+\bb_i^T\bW+K_i, \quad\quad i=1,\cdots,p,
\end{equation}
where $\bb_i=(b_{i1},\cdots,b_{ik})^T$, $(b_{1j},\cdots,b_{pj})^T=\sqrt{\lambda_j}\bgamma_j$, the factors are $\bW=(W_1,\cdots,W_k)^T$ $\sim N_k(0,\bI_k)$ and the random errors are $ (K_1,\cdots,K_p)^T$ $\sim N(0,\bA)$. Furthermore, $W_1,\cdots,W_k$ are independent of each other and independent of $K_1,\cdots,K_p$.  Changing a probability if necessary, we can assume (\ref{b20}) is the data generation process. In expression (\ref{b20}), $\{\mu_i=0\}$ correspond to the true null hypotheses, while $\{\mu_i\neq0\}$ correspond to the false ones. Note that although (\ref{b20}) is not exactly a classical multifactor model because of the existence of dependence among $K_1,\cdots,K_p$, we can nevertheless show that $(K_1,\cdots,K_p)^T$ is a weakly dependent vector if the number of factors $k$ is appropriately chosen.

We now discuss how to choose $k$ such that $(K_1,\cdots,K_p)^T$ is weakly dependent. Denote by $a_{ij}$ the $(i,j)$th element in the covariance matrix $\bA$. If we have
\begin{equation}\label{c3}
p^{-1}(\lambda_{k+1}^2+\cdots+\lambda_p^2)^{1/2} \longrightarrow 0 \ \text{as} \ p\rightarrow\infty,
\end{equation}
then by the Cauchy-Schwartz inequality
\begin{equation*}
p^{-2}\sum_{i,j}|a_{ij}|\leq p^{-1}\|\bA\|_F=p^{-1}(\lambda_{k+1}^2+\cdots+\lambda_p^2)^{1/2}\longrightarrow0 \ \text{as} \ p\rightarrow\infty.
\end{equation*}
Note that $\sum_{i=1}^p\lambda_i=tr(\bSigma)=p$, so that (\ref{c3}) is self-normalized. Note also that the left hand side of (\ref{c3}) is bounded by $p^{-1/2}\lambda_{k+1}$ which tends to zero whenever $\lambda_{k+1}=o(p^{1/2})$. Therefore, we can assume that the $\lambda_k>cp^{1/2}$ for some $c>0$. In particular, if $\lambda_1=o(p^{1/2})$, the matrix $\bSigma$ is weak dependent and $k=0$.  In practice, we always choose the smallest $k$ such that
\begin{equation*}
\frac{\sqrt{\lambda_{k+1}^2+\cdots+\lambda_{p}^2}}{\lambda_1+\cdots+\lambda_p} < \varepsilon
\end{equation*}
holds for a predetermined small $\varepsilon$, say, $0.01$.

\begin{theorem}
Suppose $(Z_1,\cdots,Z_p)^T\sim N((\mu_1,\cdots,\mu_p)^T,\bSigma)$. Choose an appropriate $k$ such that
\begin{equation*}
(C0) \ \ \ \ \ \ \ \ \frac{\sqrt{\lambda_{k+1}^2+\cdots+\lambda_{p}^2}}{\lambda_1+\cdots+\lambda_p}=O(p^{-\delta}) \ \ \ \text{for} \ \ \delta>0.
\end{equation*}
Let $\sqrt{\lambda_j}\bgamma_j=(b_{1j},\cdots,b_{pj})^T$ for $j=1,\cdots,k$. Then,
\textrm{
\begin{equation}\label{a50}
\lim_{p\rightarrow\infty}\Big\{\mathrm{FDP}(t)-\frac{\sum_{i\in\text{\{true null\}}}\Big[\Phi(a_i(z_{t/2}+\eta_i))+\Phi(a_i(z_{t/2}-\eta_i))\Big]}{\sum_{i=1}^p\Big[\Phi(a_i(z_{t/2}+\eta_i+\mu_i))+\Phi(a_i(z_{t/2}-\eta_i-\mu_i))\Big]}\Big\}=0 \ \ \text{a.s.},
\end{equation}
}where $a_i = (1-\sum_{h=1}^kb_{ih}^2)^{-1/2}$, $\eta_i = \bb_i^T\bW$ with $\bb_i=(b_{i1},\cdots,b_{ik})^T$ and $\bW\sim N_k(0,\bI_k)$ in (\ref{b20}), and $\Phi(\cdot)$ and $z_{t/2}=\Phi^{-1}(t/2)$ are the cumulative distribution function and the $t/2$ lower quantile of a standard normal distribution, respectively.
\end{theorem}
Note that condition (C0) implies that $K_1,\cdots,K_p$ are weakly dependent random variables, but (\ref{c3}) converges to zero at some polynomial rate of $p$.

Theorem 1 gives an asymptotic approximation for FDP$(t)$ under general dependence structure. To the best of our knowledge, it is the first result to explicitly spell out the impact of dependence. It is also closely connected with the existing results for independence case and weak dependence case. Let $b_{ih}=0$ for $i=1,\cdots,p$ and $h=1,\cdots,k$ in (\ref{b20}) and $K_1,\cdots,K_p$ be weakly dependent or independent normal random variables, then it reduces to the weak dependence case or independence case, respectively. In the above two specific cases, the numerator of (\ref{a50}) is just $p_0t$. Storey (2002) used an estimate for $p_0$, resulting an estimator of $\FDP(t)$ as $\widehat{p}_0t/R(t)$. This estimator has been shown to control the false discovery rate under independency and weak dependency. However, for general dependency, Storey's procedure will not work well because it ignores the correlation effect among the test statistics, as shown by (\ref{a50}). Further discussions for the relationship between our result and the other leading research for multiple testing under dependence are shown in Section 3.4.

The results in Theorem 1 can be better understood by some special dependence structures as follows. These specific cases are also considered by Roquain \& Villers (2010), Finner, Dickhaus \& Roters (2007) and Friguet, Kloareg \& Causeur (2009) under somewhat different settings.

\textbf{Example 1: [Equal Correlation]}  If $\bSigma$ has $\rho_{ij}=\rho\in[0,1)$ for $i\neq j$, then we can write
\begin{equation*}
Z_i=\mu_i+\sqrt{\rho}W+\sqrt{1-\rho}K_i \ \ \ i=1,\cdots,p
\end{equation*}
where $W\sim N(0,1)$, $K_i\sim N(0,1)$, and $W$ and all $K_i$'s are independent of each other. Thus, we have
\begin{equation*}
\lim_{p\rightarrow\infty}\Big[\mathrm{FDP}(t)-\frac{p_0\Big[\Phi(d(z_{t/2}+\sqrt{\rho}W))+\Phi(d(z_{t/2}-\sqrt{\rho}W))\Big]}{\sum_{i=1}^p\Big[\Phi(d(z_{t/2}+\sqrt{\rho}W+\mu_i))+\Phi(d(z_{t/2}-\sqrt{\rho}W-\mu_i))\Big]}\Big]=0 \ \ \text{a.s.},
\end{equation*}
where $d=(1-\rho)^{-1/2}$.

Note that Delattre \& Roquain (2011) studied the FDP in a particular case of equal correlation. They provided a slightly different decomposition of $\{Z_i\}_{i=1}^p$ in the proof of Lemma 3.3 where the errors $K_i$'s have a sum equal to 0. Finner, Dickhaus \& Roters (2007) in their Theorem 2.1 also shows a result similar to Theorem 1 for equal correlation case.

\textbf{Example 2: [Multifactor Model]} Consider a multifactor model:
\begin{equation}\label{c4}
Z_i=\mu_i+\eta_i+a_i^{-1}K_i, \quad\quad i=1,\cdots,p,
\end{equation}
where $\eta_i$ and $a_i$ are defined in Theorem 1 and $K_i\sim N(0,1)$ for $i=1,\cdots,p$. All the $W_h$'s and $K_i$'s are independent of each other. In this model, $W_1,\cdots,W_k$ are the $k$ common factors. By Theorem 1, expression (\ref{a50}) holds.

Note that the covariance matrix for model (\ref{c4}) is
$$
   \bSigma = \bL \bL^T + \diag(a_1^{-2}, \cdots, a_p^{-2}).
$$
When $\{a_j\}$ is not a constant, columns of $L$ are not necessarily eigenvectors of $\bSigma$.  In other words, when the principal component analysis is used, the decomposition of (\ref{b19}) can yield a different $L$ and condition (\ref{c3}) can require a different value of $k$.  In this sense, there is a subtle difference between our approach and that in Friguet, Kloareg \& Causeur (2009) when the principal component analysis is used.  Theorem 1 should be understood as a result for any decomposition (\ref{b19}) that satisfies condition (C0).  Because we use principal components as approximated factors, our procedure is called principal factor approximation. In practice, if one knows that the test statistics comes from a factor model structure, multiple testing procedure based on this factor model should be preferable. However, when such factor structure is not clear, our procedure can deal with an arbitrary covariance dependence case.

In Theorem 1, since $\FDP$ is bounded by 1, taking expectation on both sides of the equation (\ref{a50}) and by the Portmanteau lemma, we have the convergence of FDR:
\begin{corollary}
Under the assumptions in Theorem 1,
\begin{equation}\label{a51}
\lim_{p\rightarrow\infty}\Big\{\mathrm{FDR}(t)-E\Big[\frac{\sum_{i\in\text{\{true null\}}}\Big\{\Phi(a_i(z_{t/2}+\eta_i))+\Phi(a_i(z_{t/2}-\eta_i))\Big\}}{\sum_{i=1}^p\Big\{\Phi(a_i(z_{t/2}+\eta_i+\mu_i))+\Phi(a_i(z_{t/2}-\eta_i-\mu_i))\Big\}}\Big]\Big\}=0.
\end{equation}
\end{corollary}
The expectation on the right hand side of (\ref{a51}) is with respect to standard multivariate normal variables $(W_1,\cdots,W_k)^T\sim N_k(0,\bI_k)$.

The proof of Theorem 1 is based on the following result.
\begin{proposition}
Under the assumptions in Theorem 1,
\begin{eqnarray}
\lim_{p\rightarrow\infty}\Big[p^{-1}R(t)-p^{-1}\sum_{i=1}^p\Big[\Phi(a_i(z_{t/2}+\eta_i+\mu_i))+\Phi(a_i(z_{t/2}-\eta_i-\mu_i))\Big]\Big]=0 \ \ \text{a.s.}, \label{a54}\\
\lim_{p\rightarrow\infty}\Big[p_0^{-1}V(t)-p_0^{-1}\sum_{i\in\text{\{true null\}}}\Big[\Phi(a_i(z_{t/2}+\eta_i))+\Phi(a_i(z_{t/2}-\eta_i))\Big]\Big]=0 \ \ \text{a.s.}. \label{a54a}
\end{eqnarray}
\end{proposition}
The proofs of Theorem 1 and Proposition 2 are shown in the Appendix.

\subsection{Estimating Realized FDP}
In Theorem 1 and Proposition 2, the summation over the set of true null hypotheses is unknown. However, due to the high dimensionality and sparsity, both $p$ and $p_0$ are large and $p_1$ is relatively small. Therefore, we can use
\begin{equation}\label{a52}
\sum_{i=1}^p\Big[\Phi(a_i(z_{t/2}+\eta_i))+\Phi(a_i(z_{t/2}-\eta_i))\Big]
\end{equation}
as a conservative surrogate for
\begin{equation}\label{a53}
\sum_{i\in\text{\{true null\}}}\Big[\Phi(a_i(z_{t/2}+\eta_i))+\Phi(a_i(z_{t/2}-\eta_i))\Big].
\end{equation}
Since only $p_1$ extra terms are included in (\ref{a52}), the substitution is accurate enough for many applications.

Recall that $\FDP(t)=V(t)/R(t)$, in which $R(t)$ is observable and known. Thus, only the realization of $V(t)$ is unknown. The mean of $V(t)$ is $E\Big[\sum_{i\in\text{\{true null\}}}I(P_i\leq t)\Big]=p_0t$, since the $P$-values corresponding to the true null hypotheses are uniformly distributed. However, the dependence structure affect the variance of $V(t)$, which can be much larger than the binomial formula $p_0 t (1-t)$. Owen (2005) has theoretically studied the variance of the number of false discoveries. In our framework, expression (\ref{a52}) is a function of i.i.d. standard normal variables. Given $t$, the variance of (\ref{a52}) can be obtained by simulations and hence variance of $V(t)$ is approximated via (\ref{a52}). Relevant simulation studies will be presented in Section 5.

In recent years, there have been substantial interests in the realized random variable FDP itself in a given experiment, instead of controlling FDR, as we are usually concerned about the number of false discoveries given the observed sample of test statistics, rather than an average of FDP for hypothetical replications of the experiment. See Genovese \& Wasserman (2004), Meinshausen (2005), Efron (2007), Friguet et al (2009), etc. In our problem, by Proposition 2 it is known that $V(t)$ is approximately
\begin{equation}
\sum_{i=1}^p\Big[\Phi(a_i(z_{t/2}+\eta_i))+\Phi(a_i(z_{t/2}-\eta_i))\Big].
\end{equation}
Let
\begin{equation*}
\mathrm{FDP_A}(t)=\Big(\sum_{i=1}^p\Big[\Phi(a_i(z_{t/2}+\eta_i))+\Phi(a_i(z_{t/2}-\eta_i))\Big]\Big)/R(t),
\end{equation*}
if $R(t)\neq0$ and $\mathrm{FDP_A}(t)=0$ when $R(t)=0$. Given observations $z_1,\cdots,z_p$ of the test statistics $Z_1,\cdots,Z_p$, if the unobserved but realized factors $W_1,\cdots,W_k$ can be estimated by $\widehat{W}_1,\cdots,\widehat{W}_k$, then we can obtain an estimator of $\mathrm{FDP_A}(t)$ by
\begin{equation} \label{b21}
\widehat{\FDP}(t)=\min\Big(\sum_{i=1}^p\Big[\Phi(a_i(z_{t/2}+\widehat{\eta}_i))+\Phi(a_i(z_{t/2}-\widehat{\eta}_i))\Big],R(t)\Big)/R(t),
\end{equation}
when $R(t)\neq0$ and $\widehat{\FDP}(t)=0$ when $R(t)=0$. Note that in (\ref{b21}), $\widehat{\eta}_i=\sum_{h=1}^kb_{ih}\widehat{W}_h$ is an estimator for $\eta_i=\bb_i^T\bW$.

The following procedure is one practical way to estimate $\bW=(W_1,\cdots,W_k)^T$ based on the data. For observed values $z_1,\cdots,z_p$, we choose the smallest $90\%$ of $|z_i|$'s, say. For ease of notation, assume the first $m$ $z_i$'s have the smallest absolute values. Then approximately
\begin{equation} \label{b22}
Z_i=\bb_i^T\bW+K_i,\quad i=1,\cdots,m.
\end{equation}
The approximation from (\ref{b20}) to (\ref{b22}) stems from the intuition that large $|\mu_i|$'s tend to produce large $|z_i|$'s as $Z_i\sim N(\mu_i,1)$ $1\leq i\leq p$ and the sparsity makes approximation errors negligible. Finally we apply the robust $L_1$-regression to the equation set (\ref{b22}) and obtain the least-absolute deviation estimates $\widehat{W}_1,\cdots,\widehat{W}_k$. We use $L_1$-regression rather than $L_2$-regression because there might be nonzero $\mu_i$ involved in (\ref{b22}) and $L_1$ is more robust to the outliers than $L_2$. Other possible methods include using penalized method such as LASSO or SCAD to explore the sparsity.  For example, one can minimize
$$
   \sum_{i=1}^p (Z_i - \mu_i - \bb_i^T\bW)^2 + \sum_{i=1}^p p_\lambda(|\mu_i|)
$$
with respect to $\{\mu_i\}_{i=1}^p$ and $\bW$, where $p_\lambda(\cdot)$ is a folded-concave penalty function (Fan and Li, 2001).

The estimator (\ref{b21}) performs significantly better than Efron (2007)'s estimator in our simulation studies. One difference is that in our setting $\bSigma$ is known. The other is that we give a better approximation as shown in Section 3.4.

Efron (2007) proposed the concept of conditional FDR. Consider $E(V(t))/R(t)$ as one type of FDR definitions (see Efron (2007) expression (46)). The numerator $E(V(t))$ is over replications of the experiment, and equals a constant $p_0t$. But if the actual correlation structure in a given experiment is taken into consideration, then Efron (2007) defines the conditional FDR as $E(V(t)|A)/R(t)$ where $A$ is a random variable which measures the dependency information of the test statistics. Estimating the realized value of $A$ in a given experiment by $\widehat{A}$, one can have the estimated conditional FDR as $E(V(t)|\widehat{A})/R(t)$. Following Efron's proposition, Friguet et al (2009) gave the estimated conditional FDR by $E(V(t)|\widehat{\bW})/R(t)$ where $\widehat{\bW}$ is an estimate of the realized random factors $\bW$ in a given experiment.

Our estimator in (\ref{b21}) for the realized FDP in a given experiment can be understood as an estimate of conditional FDR. Note that (\ref{a53}) is actually $E(V(t)|\{\eta_i\}_{i=1}^p)$. By Proposition 2, we can approximate $V(t)$ by $E(V(t)|\{\eta_i\}_{i=1}^p)$. Thus the estimate of conditional FDR $E(V(t)|\{\widehat{\eta}_i\}_{i=1}^p)/R(t)$ is directly an estimate of the realized FDP $V(t)/R(t)$ in a given experiment.

\subsection{Asymptotic Justification}
Let $\bw=(w_1,\cdots,w_k)^T$ be the realized values of $\{W_h\}_{h=1}^k$, and $\widehat{\bw}$ be an estimator for $\bw$. We now show in Theorem 2 that $\widehat{\FDP}(t)$ in (\ref{b21}) based on a consistent estimator $\widehat{\bw}$ has the same convergence rate as $\widehat{\bw}$ under some mild conditions.
\begin{theorem}
If the following conditions are satisfied:
\begin{itemize}
\item[(C1)] $R(t)/p>H$ for $H>0$ as $p\rightarrow\infty$,
\item[(C2)] $\min_{1\leq i\leq p}\min(|z_{t/2}+\bb_i^T\bw|,|z_{t/2}-\bb_i^T\bw|)\geq \tau>0$,
\item[(C3)] $\|\widehat{\bw}-\bw\|_2=O_p(p^{-r})$ for some $r>0$,
\end{itemize}
then $|\mathrm{\widehat{FDP}}(t)-\mathrm{FDP_A}(t)|=O(\|\hw-\bw\|_2)$.
\end{theorem}
In Theorem 2, (C2) is a reasonable condition because $z_{t/2}$ is a large negative number when threshold $t$ is small and $\bb_i^T\bw$ is a realization from a normal distribution $N(0, \sum_{h=1}^kb_{ih}^2)$ with $\sum_{h=1}^kb_{ih}^2<1$. Thus $z_{t/2}+\bb_i^T\bw$ or $z_{t/2}-\bb_i^T\bw$ is unlikely close to zero.

Theorem 3 shows the asymptotic consistency of $L_1-$regression estimators under model (\ref{b22}). Portnoy (1984b) has proven the asymptotic consistency for robust regression estimation when the random errors are i.i.d. However, his proof does not work here because of the weak dependence of  random errors. Our result allows $k$ to grow with $m$, even at a faster rate of $o(m^{1/4})$ imposed by Portnoy (1984b).
\begin{theorem}
Suppose (\ref{b22}) is a correct model. Let $\hw$ be the $L_1-$regression estimator:
\begin{equation}
\hw \equiv \argmin_{\bbeta\in R^k}\sum_{i=1}^m|Z_i-\bb_i^T\bbeta|
\end{equation}
where $\bb_i=(b_{i1},\cdots,b_{ik})^T$. Let $\bw=(w_1,\cdots,w_k)^T$ be the realized values of $\{W_h\}_{h=1}^k$.
Suppose $k=O(m^{\kappa})$ for $0\leq\kappa<1-\delta$. Under the assumptions
\begin{itemize}
\item[(C4)]\label{a55}
$\sum_{j=k+1}^p\lambda_j^2\leq\eta$ for $\eta=O(m^{2\kappa})$,
\item[(C5)]\label{a56}
\begin{equation*}
\lim_{m\rightarrow\infty}\sup_{\|\bu\|=1}m^{-1}\sum_{i=1}^mI(|\bb_i^T\bu|\leq d)=0
\end{equation*}
for a constant $d>0$.
\item[(C6)] $a_{\max}/a_{\min}\leq S$ for some constant $S$ when $m\rightarrow\infty$ where $1/a_i$ is the standard deviation of $K_i$,
\item[(C7)] $a_{\min}=O(m^{(1-\kappa)/2})$.
\end{itemize}
We have $\|\hw-\bw\|_2=O_p(\sqrt{\frac{k}{m}})$.
\end{theorem}
(C4) is stronger than (C0) in Theorem 1 as (C0) only requires $\sum_{j=k+1}^p\lambda_j^2=O(p^{2-2\delta})$. (C5) ensures the identifiability of $\bbeta$, which is similar to Proposition 3.3 in Portnoy (1984a). (C6) and (C7) are imposed to facilitate the technical proof.

We now briefly discuss the role of the factor $k$.  To make the approximation in Theorem 1 hold, we need $k$ to be large.  On the other hand, to make the realized factors estimable with reasonably accuracy, we hope to choose a small $k$ as demonstrated in Theorem 3.  Thus, the practical choice of $k$ should be done with care.

Since $m$ is chosen as a certain large proportion of $p$, combination of Theorem 2 and Theorem 3 thus shows the asymptotic consistency of $\widehat{\FDP}(t)$ based on $L_1-$regression estimator of $\bw=(w_1,\cdots,w_k)^T$ in model (\ref{b22}):
\begin{equation*}
|\mathrm{\widehat{FDP}}(t)-\mathrm{FDP_A}(t)|=O_p(\sqrt{\frac{k}{m}}).
\end{equation*}

\indent The result in Theorem 3 are based on the assumption that (\ref{b22}) is a correct model. In the following, we will show that even if (\ref{b22}) is not a correct model, the effects of misspecification are negligible when $p$ is sufficiently large. To facilitate the mathematical derivations, we instead consider the least-squares estimator. Suppose we are estimating $\bW=(W_1,\cdots,W_k)^T$ from (\ref{b20}). Let $\bX$ be the design matrix of model (\ref{b20}). Then the least-squares estimator for $\bW$ is $\widehat{\bW}_{\LS}=\bW+(\bX^T\bX)^{-1}\bX^T(\bmu+\bK)$, where $\bmu=(\mu_1,\cdots,\mu_p)^T$ and $\bK=(K_1,\cdots,K_p)^T$. Instead, we estimate $W_1,\cdots,W_k$ based on the simplified model (\ref{b22}), which ignores sparse $\{\mu_i\}$. Then the least-squares estimator for $\bW$ is $\widehat{\bW}_{\LS}^*=\bW+(\bX^T\bX)^{-1}\bX^T\bK=\bW$, in which we utilize the orthogonality between $\bX$ and $\var(\bK)$. The following result shows that the effect of misspecification in model (\ref{b22}) is negligible when $p\rightarrow\infty$, and consistency of the least-squares estimator.
\begin{theorem}
The bias due to ignoring non-nulls is controlled by
\begin{equation*}
\|\widehat{\bW}_{\mathrm{LS}}-\bW\|_2=\|\widehat{\bW}_{\mathrm{LS}}-\widehat{\bW}_{\mathrm{LS}}^*\|_2\leq\|\bmu\|_2\Big(\sum_{i=1}^k\lambda_i^{-1}\Big)^{1/2}.
\end{equation*}
\end{theorem}

In Theorem 1, we can choose appropriate $k$ such that $\lambda_k>cp^{1/2}$ as noted proceeding Theorem 1. Therefore, $\sum_{i=1}^k\lambda_i^{-1}\rightarrow 0$ as $p\rightarrow\infty$ is a reasonable condition. When $\{\mu_i\}_{i=1}^p$ are truly sparse, it is expected that $\|\bmu\|_2$ grows slowly or is even bounded so that the bound in Theorem 4 is small. For $L_1-$regression, it is expected to be even more robust to the outliers in the sparse vector $\{\mu_i\}_{i=1}^p$.

\subsection{Dependence-Adjusted Procedure}

A problem of the method used so far is that the ranking of statistical significance is completely determined by the ranking of the test statistics $\{|Z_i|\}$.  This is undesirable and can be inefficient for the dependent case: the correlation structure should also be taken into account. We now show how to use the correlation structure to improve the signal to noise ratio.

Note that by (\ref{b20}), $Z_i-\bb_i^T\bW\sim N(\mu_i,a_i^{-2})$ where $a_i$ is defined in Theorem 1. Since $a_i^{-1}\leq1$, the signal to noise ratio increases, which makes the resulting procedure more powerful. Thus, if we know the true values of the common factors $\bW=(W_1,\cdots,W_k)^T$, we can use $a_i(Z_i-\bb_i^T\bW)$ as the test statistics. The dependence-adjusted $p$-values $\displaystyle 2\Phi(-|a_i(Z_i-\bb_i^T\bW)|)$ can then be used. Note that this testing procedure has different thresholds for different hypotheses based on the magnitude of $Z_i$, and has incorporated the correlation information among hypotheses. In practice, given $Z_i$, the common factors $\{W_h\}_{h=1}^k$ are realized but unobservable. As shown in section 3.2, they can be estimated. The dependence adjusted $p$-values are then given by
\begin{equation}\label{d1}
2\Phi(-|a_i(Z_i-\bb_i^T\widehat{\bW})|)
\end{equation}
for ranking the hypotheses where $\widehat{\bW}=(\widehat{W}_1,\cdots,\widehat{W}_k)^T$ is an estimate of the principal factors. We will show in section 5 by simulation studies that this dependence-adjusted procedure is more powerful.  The ``factor adjusted multiple testing procedure" in Friguet et al (2009) shares a similar idea.

\subsection{Relation with Other Methods}
Efron (2007) proposed a novel parametric model for $V(t)$:
\begin{equation}
V(t)=p_0t\Big[1+2A\frac{(-z_{t/2})\phi(z_{t/2})}{\sqrt{2}t}\Big],
\end{equation}
where $A\sim N(0,\alpha^2)$ for some real number $\alpha$ and $\phi(\cdot)$ stands for the probability density function of the standard normal distribution. The correlation effect is explained by the dispersion variate $A$. His procedure is to estimate $A$ from the data and use
\begin{equation}
p_0t\Big[1+2\widehat{A}\frac{(-z_{t/2})\phi(z_{t/2})}{\sqrt{2}t}\Big]\Big/R(t)
\end{equation}
as an estimator for realized $\FDP(t)$. Note that the above expressions are adaptations from his procedure for the one-sided test to our two-sided test setting. In his simulation, the above estimator captures the general trend of the FDP, but it is not accurate and deviates from the true FDP with large amount of variability. Consider our estimator $\widehat{\mathrm{FDP}}(t)$ in (\ref{b21}). Write $\widehat{\eta}_i=\sigma_iQ_i$ where $Q_i\sim N(0,1)$. When $\sigma_i\rightarrow0$ for $\forall i\in\{\text{true null}\}$, by the second order Taylor expansion,

\begin{equation}
\widehat{\mathrm{FDP}}(t)\approx\frac{p_0t}{R(t)}\Big[1+\sum_{i\in\{\text{true null}\}}\sigma_i^2(Q_i^2-1)\frac{(-z_{t/2})\phi(z_{t/2})}{p_0t}\Big].
\end{equation}
By comparison with Efron's estimator, we can see that
\begin{equation}
\widehat{A}=\frac{1}{\sqrt{2}p_0}\sum_{i\in\{\text{true null}\}}\Big[\widehat{\eta}_i^2-E(\widehat{\eta}_i^2)\Big].
\end{equation}
Thus, our method is more general.

Leek \& Storey (2008) considered a general framework for modeling the dependence in multiple testing. Their idea is to model the dependence via a factor model and reduces the multiple testing problem from dependence to independence case via accounting the effects of common factors. They also provided a method of estimating the common factors. In contrast, our problem is different from Leek \& Storey's and we estimate common factors from principal factor approximation and other methods. In addition, we provide the approximated FDP formula and its consistent estimate.

Friguet, Kloareg \& Causeur (2009) followed closely the framework of Leek \& Storey (2008). They assumed that the data come directly from a multifactor model with independent random errors, and then used the EM algorithm to estimate all the parameters in the model and obtained an estimator for FDP$(t)$. In particular, they subtract $\eta_i$ out of (\ref{c4}) based on their estimate from the EM algorithm to improve the efficiency. However, the ratio of estimated number of factors to the true number of factors in their studies varies according to the dependence structures by their EM algorithm, thus leading to inaccurate estimated $\FDP(t)$. Moreover, it is hard to derive theoretical results based on the estimator from their EM algorithm. Compared with their results, our procedure does not assume any specific dependence structure of the test statistics. What we do is to decompose the test statistics into an approximate factor model with weakly dependent errors, derive the factor loadings and estimate the unobserved but realized factors by $L_1$-regression. Since the theoretical distribution of $V(t)$ is known, estimator (\ref{b21}) performs well based on a good estimation for $W_1,\cdots,W_k$.

\section{Approximate Estimation of FDR}
In this section we propose some ideas that can asymptotically control the FDR, not the FDP, under arbitrary dependency. Although their validity is yet to be established, promising results reveal in the simulation studies. Therefore, they are worth some discussion and serve as a direction of our future work.

Suppose that the number of false null hypotheses $p_1$ is known. If the signal $\mu_i$ for $i\in\text{\{false null\}}$ is strong enough such that
\begin{equation} \label{b24}
\Phi\Big(a_i(z_{t/2}+\eta_i+\mu_i)\Big)+\Phi\Big(a_i(z_{t/2}-\eta_i-\mu_i)\Big)\approx1,
\end{equation}
then asymptotically the FDR is approximately given by
\begin{equation} \label{b25}
\FDR(t)=E\Big\{\frac{\sum_{i=1}^p\Big[\Phi(a_i(z_{t/2}+\eta_i))+\Phi(a_i(z_{t/2}-\eta_i))\Big]}{\sum_{i=1}^p\Big[\Phi(a_i(z_{t/2}+\eta_i))+\Phi(a_i(z_{t/2}-\eta_i))\Big]+p_1}\Big\},
\end{equation}
which is the expectation of a function of $W_1,\cdots,W_k$. Note that $\FDR(t)$ is a known function and can be computed by Monte Carlo simulation. For any predetermined error rate $\alpha$, we can use the bisection method to solve $t$ so that $\FDR(t)=\alpha$. Since $k$ is not large, the Monte Carlo computation is sufficiently fast for most applications.

The requirement (\ref{b24}) is not very strong. First of all, $\Phi(3)\approx0.9987$, so (\ref{b24}) will hold if any number inside the $\Phi(\cdot)$ is greater than 3. Secondly, $1-\sum_{h=1}^kb_{ih}^2$ is usually very small. For example, if it is $0.01$, then $a_i=(1-\sum_{h=1}^kb_{ih}^2)^{-1/2}\approx10$, which means that if either $z_{t/2}+\eta_i+\mu_i$ or $z_{t/2}-\eta_i-\mu_i$ exceed 0.3, then (\ref{b24}) is approximately satisfied. Since the effect of sample size $n$ is involved in the problem in Section 2, (\ref{b24}) is not a very strong condition on the signal strength $\{\beta_i\}$.

Note that Finner et al (2007) considered a ``Dirac uniform model", where the $p$-values corresponding to a false hypothesis are exactly equal to 0. This model might be potentially useful for FDR control. The calculation of (\ref{b25}) requires the knowledge of the proportion $p_1$ of signal in the data. Since $p_1$ is usually unknown in practice, there is also future research interest in estimating $p_1$ under arbitrary dependency.

\section{Simulation Studies}
In the simulation studies, we consider $p=2000$, $n=100$, $\sigma=2$, the number of false null hypotheses $p_1=10$ and the nonzero $\beta_i=1$, unless stated otherwise. We will present 6 different dependence structures for $\bSigma$ of the test statistics $(Z_1,\cdots,Z_p)^T\sim N((\mu_1,\cdots,\mu_p)^T,\bSigma)$. Following the setting in section 2, $\bSigma$ is the correlation matrix of a random sample of size $n$ of $p-$dimensional vector $\bX_i=(X_{i1},\cdots,X_{ip})$, and $\mu_j=\sqrt{n}\beta_j\widehat{\sigma}_j/\sigma$, $j=1,\cdots,p$. The data generating process vector $\bX_i$'s are as follows.
\begin{itemize}
\item \textbf{[Equal correlation]} Let $\bX^T=(X_{1},\cdots,X_{p})^T\sim N_p(0,\bSigma)$ where $\bSigma$ has diagonal element 1 and off-diagonal element $1/2$.
\item \textbf{[Fan \& Song's model]} For $\bX=(X_{1},\cdots,X_{p})$, let $\{X_{k}\}_{k=1}^{1900}$ be i.i.d. $N(0,1)$ and
    \begin{equation*}
    X_{k}=\sum_{l=1}^{10}X_{l}(-1)^{l+1}/5+\sqrt{1-\frac{10}{25}}\epsilon_{k}, \ \ k=1901,\cdots,2000,
    \end{equation*}
    where $\{\epsilon_{k}\}_{k=1901}^{2000}$ are standard normally distributed.
\item \textbf{[Independent Cauchy]} For $\bX=(X_{1},\cdots,X_{p})$, let $\{X_{k}\}_{k=1}^{2000}$ be i.i.d. Cauchy random variables with location parameter 0 and scale parameter 1.
\item \textbf{[Three factor model]} For $\bX=(X_{1},\cdots,X_{p})$, let
    \begin{equation*}
    X_{j}=\rho_j^{(1)}W^{(1)}+\rho_j^{(2)}W^{(2)}+\rho_j^{(3)}W^{(3)}+H_{j},
    \end{equation*}
    where $W^{(1)}\sim N(-2,1)$, $W^{(2)}\sim N(1,1)$, $W^{(3)}\sim N(4,1)$, $\rho_{j}^{(1)},\rho_{j}^{(2)},\rho_{j}^{(3)}$ are i.i.d. $U(-1,1)$, and $H_{j}$ are i.i.d. $N(0,1)$.
\item \textbf{[Two factor model]} For $\bX=(X_{1},\cdots,X_{p})$, let
    \begin{equation*}
    X_{j}=\rho_j^{(1)}W^{(1)}+\rho_j^{(2)}W^{(2)}+H_{j},
    \end{equation*}
    where $W^{(1)}$ and $W^{(2)}$ are i.i.d. $N(0,1)$, $\rho_{j}^{(1)}$ and $\rho_{j}^{(2)}$ are i.i.d. $U(-1,1)$, and $H_{j}$ are i.i.d. $N(0,1)$.
\item \textbf{[Nonlinear factor model]} For $\bX=(X_{1},\cdots,X_{p})$, let
    \begin{equation*}
    X_{j}=\sin(\rho_j^{(1)}W^{(1)})+sgn(\rho_j^{(2)})\exp(|\rho_j^{(2)}|W^{(2)})+H_{j},
    \end{equation*}
    where $W^{(1)}$ and $W^{(2)}$ are i.i.d. $N(0,1)$, $\rho_{j}^{(1)}$ and $\rho_{j}^{(2)}$ are i.i.d. $U(-1,1)$, and $H_{j}$ are i.i.d. $N(0,1)$.
\end{itemize}

Fan \& Song's Model has been considered in Fan \& Song (2010) for high dimensional variable selection. This model is close to the independent case but has some special dependence structure. Note that although we have used the term ``factor model" above to describe the dependence structure, it is not the factor model for the test statistics $Z_1,\cdots,Z_p$ directly. The covariance matrix of these test statistics is the sample correlation matrix of $X_1,\cdots,X_p$. 

The effectiveness of our method is examined in several aspects.  We first examine the goodness of approximation in Theorem 1 by comparing the marginal distributions and variances.  We then compare the accuracy of FDP estimates with other methods.  Finally, we demonstrate the improvement of the power with dependence adjustment.

\begin{figure}[h!!!]
\setlength{\unitlength}{1mm}
\begin{center}
\scalebox{0.46}{\includegraphics{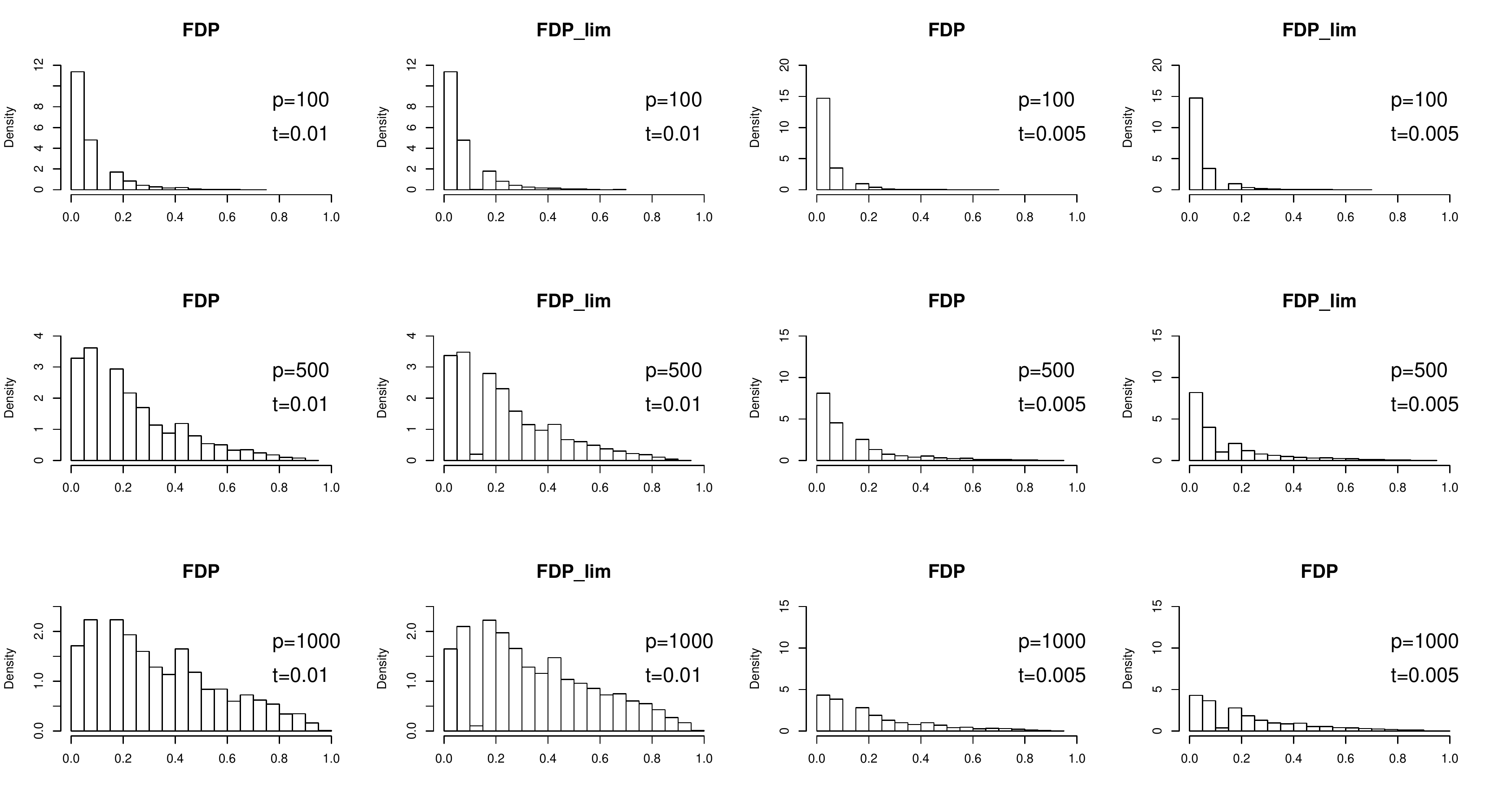}}
\end{center}
\caption{Comparison for the distribution of the FDP with that of its approximated expression, based on the two factor model over 10000 simulations. From the top row to the bottom, $p=100, 500, 1000$ respectively. The first two columns correspond to $t=0.01$ and the last two correspond to $t=0.005$. 
}
\label{ar}
\label{a59}
\end{figure}

\textbf{Distributions of FDP and its approximation:}  Without loss of generality, we consider a dependence structure based on the two factor model above. Let $n=100$, $p_1=10$ and $\sigma=2$. Let $p$ vary from 100 to 1000 and $t$ be either 0.01 or 0.005.  The distributions of $\FDP(t)$ and its approximated expression in Theorem 1 are plotted in Figure~\ref{a59}.  The convergence of the distributions are self-evidenced. Table 2 summarizes the total variation distance between the two distributions.

\begin{table}[h!!!]
\begin{center}\caption{Total variation distance between the distribution of $\FDP$ and the limiting distribution of $\FDP$ in Figure 1. The total variation distance is calculated based on ``TotalVarDist" function with ``smooth" option in R software.}
  \begin{tabular}{rccc}
  \hline\hline
        & $p=100$ &$p=500$ &$p=1000$\\
  \hline
  $t=0.01$ & 0.6668     &0.1455        & 0.0679       \\
  $t=0.005$ & 0.6906    &0.2792        & 0.1862       \\
  \hline
  \end{tabular}
\end{center}
\end{table}

\textbf{Variance of $V(t)$:} Variance of false discoveries in the correlated test statistics is usually large compared with that of the independent case which is $p_0t(1-t)$, due to correlation structures. Thus the ratio of variance of false discoveries in the dependent case to that in the independent test statistics can be considered as a measure of correlation effect. See Owen (2005). Estimating the variance of false discoveries is an interesting problem.  With approximation (\ref{a54a}),  this can easily be computed. In Table~\ref{e0}, we compare the true variance of the number of false discoveries, the variance of expression (\ref{a53}) (which is infeasible in practice) and the variance of expression (\ref{a52}) under 6 different dependence structures. It shows that the variance computed based on expression (\ref{a52}) approximately equals the variance of number of false discoveries. Therefore, we provide a fast and alternative method to estimate the variance of number of false discoveries in addition to the results in Owen (2005). Note that the variance for independent case is merely less than 2. The impact of dependence is very substantial.

\begin{table}[h!!!]
\begin{center}\caption{Comparison for variance of number of false discoveries (column 2), variance of expression (\ref{a53}) (column 3) and variance of expression (\ref{a52}) (column 4) with $t=0.001$ based on 10000 simulations. }\label{e0}
  \begin{tabular}{crrr}
  \hline\hline
   Dependence Structure  & $\var(V(t))$  & $\var(V)$  & $\var(V.up)$\\
   \hline
     Equal correlation   &  180.9673     &  178.5939  & 180.6155    \\
     Fan \& Song's model &   5.2487     & 5.2032   &   5.2461  \\
     Independent Cauchy  &  9.0846     & 8.8182   &   8.9316  \\
     Three factor model  &   81.1915   & 81.9373    & 83.0818  \\
     Two factor model    &  53.9515    & 53.6883    &  54.0297   \\
     Nonlinear factor model  &  48.3414    & 48.7013    &   49.1645  \\
   \hline
   \end{tabular}
\end{center}
\end{table}

\begin{table}[h!!!]
\begin{center}\caption{Comparison of FDP values for our method based on equation (\ref{b25}) without taking expectation (PFA) with Storey's procedure and Benjamini-Hochberg's procedure under six different dependence structures, where $p=2000$, $n=200$, $t=0.001$, and $\beta_i=1$ for $i\in\text{\{false null\}}$. The computation is based on 10000 simulations. The means of FDP are listed with the standard deviations in the brackets.}\label{e1}
  \begin{tabular}{ccccc}
  \hline\hline
                         & True FDP  & PFA  & Storey   & B-H\\
   \hline
     Equal correlation   & $6.67\%$     & $6.61\%$   & $2.99\%$  &  $3.90\%$   \\
                         & ($15.87\%$)  &($15.88\%$) &($10.53\%$)& ($14.58\%$) \\
     Fan \& Song's model & $14.85\%$     & $14.85\%$    & $13.27\%$   & $14.46\%$   \\
                         & ($11.76\%$)   &($11.58\%$)   &($11.21\%$)  &($13.46\%$)  \\
     Independent Cauchy  & $13.85\%$     & $13.62\%$  & $11.48\%$  & $13.21\%$  \\
                         &($13.60\%$)    &($13.15\%$) &($12.39\%$) &($15.40\%$) \\
     Three factor model  & $8.08\%$      & $8.29\%$    & $4.00\%$   & $5.46\%$  \\
                         &($16.31\%$)   &($16.39\%$)   & ($11.10\%$)& ($16.10\%$)\\
     Two factor model    & $8.62\%$     & $8.50\%$     & $4.70\%$ & $5.87\%$    \\
                         &($16.44\%$)   &($16.27\%$)   &($11.97\%$) &($16.55\%$) \\
     Nonlinear factor model  & $6.63\%$     & $6.81\%$     & $3.20\%$ & $4.19\%$    \\
                         & ($15.56\%$)    & ($15.94\%$)    &($10.91\%$) &($15.31\%$)\\
   \hline
   \end{tabular}
\end{center}
\end{table}

\textbf{Comparing methods of estimating FDP:} Under different dependence structures, we compare FDP values using our procedure PFA in equation (\ref{b25}) without taking expectation and with $p_1$ known, Storey's procedure with $p_1$ known ($(1-p_1)t/R(t)$) and Benjamini-Hochberg procedure. Note that Benjamini-Hochberg procedure is a FDR control procedure rather than a FDP estimating procedure. The Benjamini-Hochberg FDP is obtained by using the mean of ``True FDP" in Table~\ref{e1} as the control rate in B-H procedure. Table~\ref{e1} shows that our method performs much better than Storey's procedure and Benjamini-Hochberg procedure, especially under strong dependence structures (rows 1, 4, 5, and 6), in terms of both mean and variance of the distribution of FDP. Recall that the expected value of FDP is the FDR. Table 3 also compares the FDR of three procedures by looking at the averages. Note that the actual FDR from B-H procedure under dependence is much smaller than the control rate, which suggests that B-H procedure can be quite conservative under dependence.

\begin{figure}[h!!!]
\begin{center}
\scalebox{0.46}{\includegraphics{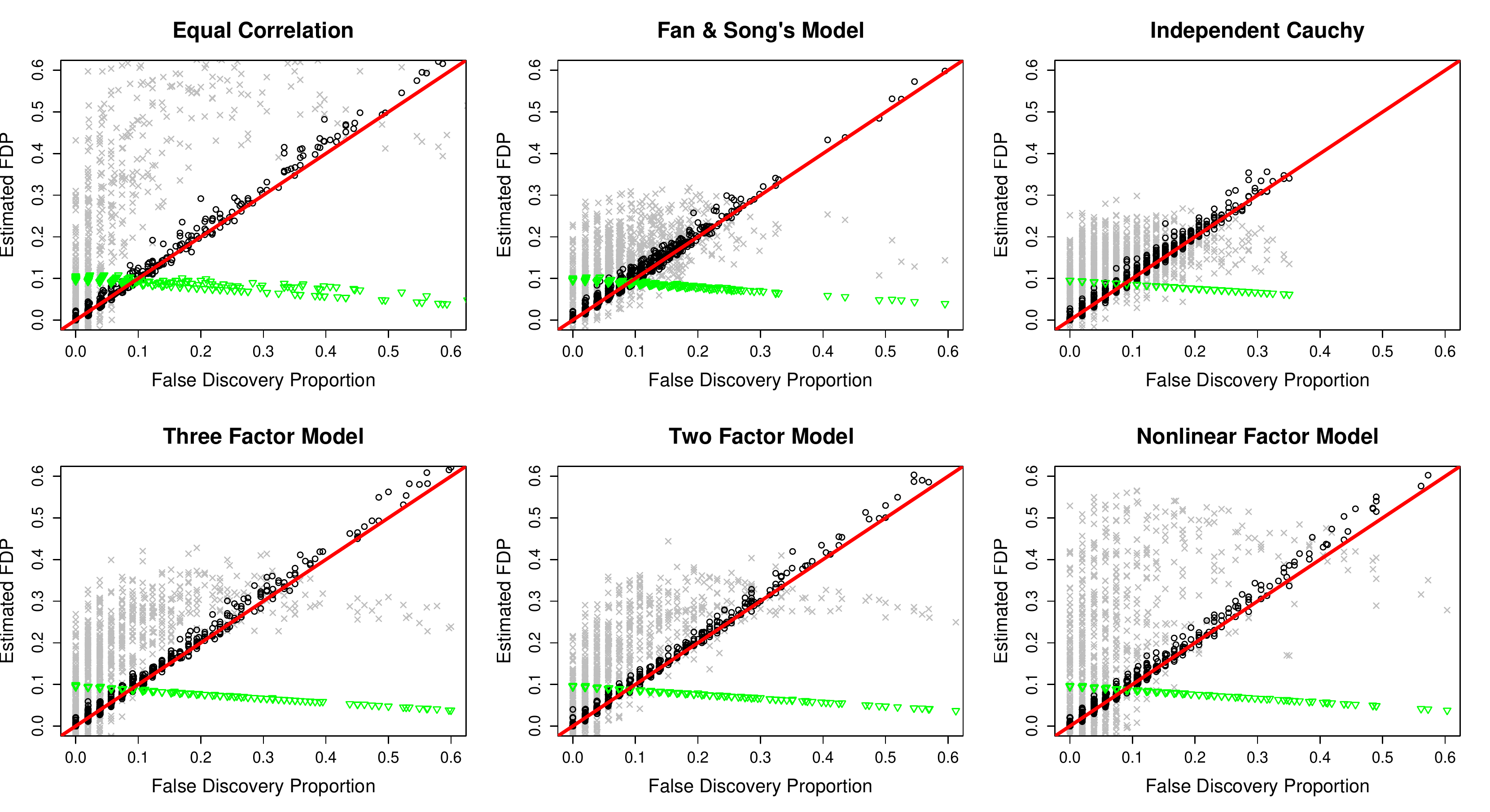}}
\end{center}
\vspace{-0.3cm}
\caption{Comparison of true values of False Discovery Proportion with estimated FDP by Efron (2007)'s procedure (grey crosses) and our PFA method (black dots) under six different dependence structures, with $p=1000$, $p_1=50$, $n=100$, $\sigma=2$, $t=0.005$ and $\beta_i=1$ for $i\in\text{\{false null\}}$ based on 1000 simulations. The $Z$-statistics with absolute value less than or equal to $x_0=1$ are used to estimate the dispersion variate $A$ in Efron (2007)'s estimator. The unconditional estimate of $\FDR(t)$ is $p_0t/R(t)$ shown as green triangles.}
\label{a60}
\end{figure}

\begin{figure}[h!!!]
\begin{center}
\scalebox{0.45}{\includegraphics{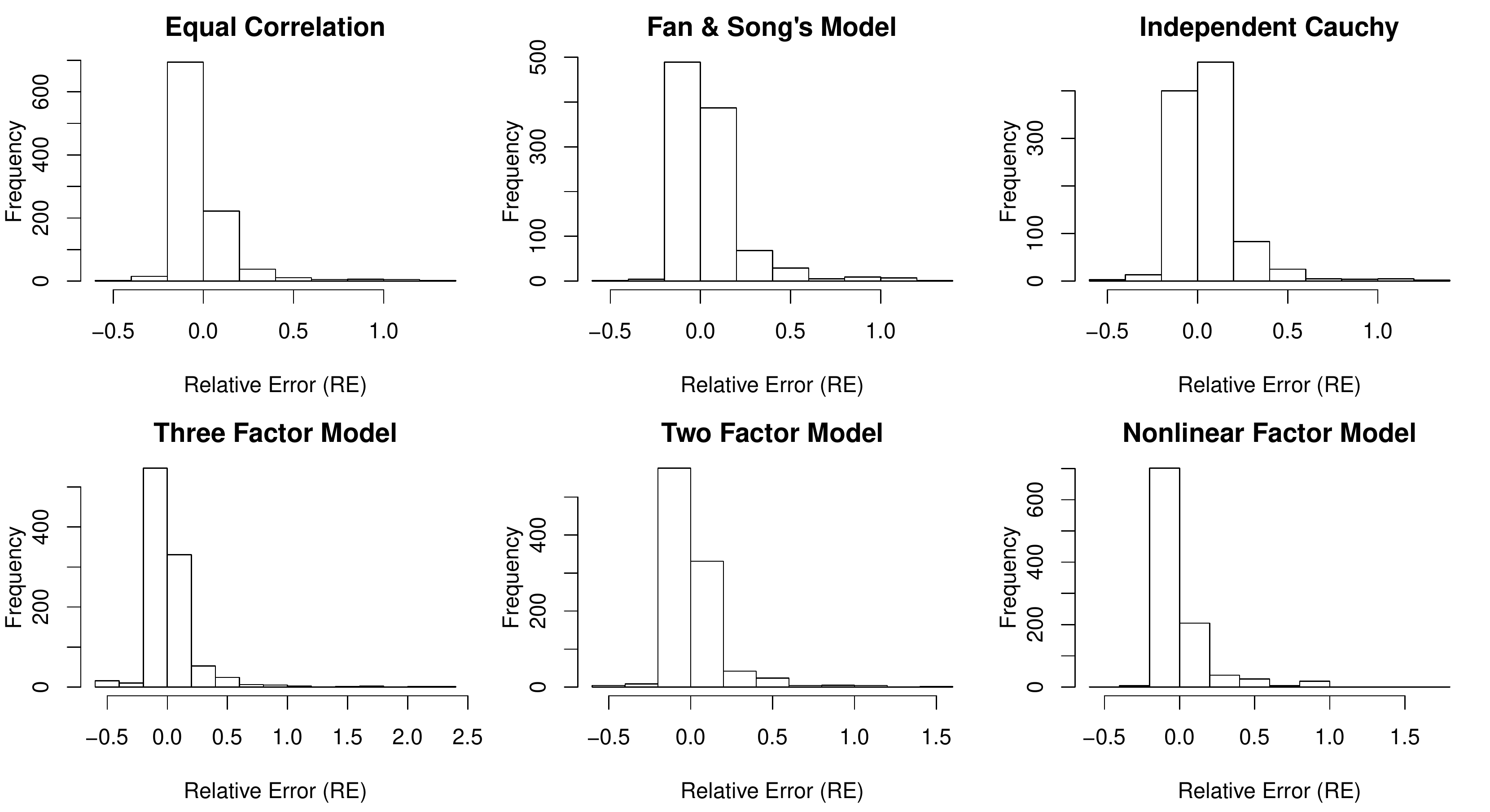}}
\end{center}
\vspace{-0.3cm}
\caption{Histograms of the relative error (RE) between true values of FDP and estimated FDP by our PFA method under the six dependence structures in Figure~\ref{a60}. RE is defined as $(\widehat{\text{FDP}}(t)-\text{FDP}(t))/\text{FDP}(t)$ if $\text{FDP}(t) \neq 0$ and 0 otherwise.}\label{c7}
\end{figure}

\textbf{Comparison with Efron's Methods:} We now compare the estimated values of our method PFA (\ref{b21}) and Efron (2007)'s estimator with true values of false discovery proportion, under 6 different dependence structures. Efron (2007)'s estimator is developed for estimating FDP under unknown $\bSigma$. In our simulation study, we have used a known $\Sigma$ for Efron's estimator for fair comparisons. The results are depicted in Figure~\ref{a60}, Figure~\ref{c7} and Table~\ref{e2}. Figure~\ref{a60} shows that our estimated values correctly track the trends of FDP with smaller amount of noise. It also shows that both our estimator and Efron's estimator tend to overestimate the true FDP, since $\mathrm{FDP_A}(t)$ is an upper bound of the true $\mathrm{FDP}(t)$. They are close only when the number of false nulls $p_1$ is very small. In the current simulation setting, we choose $p_1=50$ compared with $p=1000$, therefore, it is not a very sparse case. However, even under this case, our estimator still performs very well for six different dependence structures. Efron (2007)'s estimator is illustrated in Figure~\ref{a60} with his suggestions for estimating parameters, which captures the general trend of true FDP but with large amount of noise. Figure~\ref{c7} shows that the relative errors of PFA concentrate around 0, which suggests good accuracy of our method in estimating FDP. Table~\ref{e2} summarizes the relative errors of the two methods.


\begin{table}[h!]
\begin{center}\caption{Means and standard deviations of the relative error between true values of FDP and estimated FDP under the six dependence structures in Figure~\ref{a60}. $\text{RE}_{\text{P}}$ and $\text{RE}_{\text{E}}$ are the relative errors of our PFA estimator and Efron (2007)'s estimator, respectively. RE is defined in Figure~\ref{c7}.}\label{e2}
  \begin{tabular}{c|cc|cc}
  \hline\hline
                   &\multicolumn{2}{c|}{$\text{RE}_{\text{P}}$} &\multicolumn{2}{c}{$\text{RE}_{\text{E}}$}\\
                    \cline{2-3} \cline{4-5}
                   & mean  & SD   & mean  & SD\\
   \hline
     Equal correlation   & 0.0241   & 0.1262   & 1.4841     &3.6736  \\
     Fan \& Song's model & 0.0689   & 0.1939   & 1.2521     &1.9632  \\
     Independent Cauchy  & 0.0594   & 0.1736   & 1.3066     &2.1864  \\
     Three factor model  & 0.0421   & 0.1657   & 1.4504     &2.6937  \\
     Two factor model    & 0.0397   & 0.1323   & 1.1227     &2.0912  \\
     Nonlinear factor model     &  0.0433   &  0.1648  &  1.3134    &4.0254 \\
   \hline
   \end{tabular}
\end{center}
\end{table}

\textbf{Dependence-Adjusted Procedure:} We compare the dependence-adjusted procedure described in section 3.4 with the testing procedure based only on the observed test statistics without using correlation information.  The latter is to compare the original z-statistics with a fixed threshold value and is labeled as ``fixed threshold procedure'' in Table~\ref{e5}. With the same FDR level, a procedure with smaller false nondiscovery rate (FNR) is more powerful, where $\FNR=E[T/(p-R)]$ using the notation in Table 1.
\begin{table}[h!!!]
\begin{center}\caption{Comparison of Dependence-Adjusted Procedure with Fixed Threshold Procedure under six different dependence structures, where $p=1000$, $n=100$, $\sigma=1$, $p_1=200$, nonzero $\beta_i$ simulated from $U(0,1)$ and $k=n-3$ over 1000 simulations.}\label{e5}
\begin{tabular}{c|c|c|c|c|c|c}
  \hline\hline
         &\multicolumn{3}{c|}{Fixed Threshold Procedure} &\multicolumn{3}{c}{Dependence-Adjusted Procedure}\\
  \cline{2-4} \cline{5-7}
         &FDR   &FNR   &Threshold  & FDR  &FNR  &Threshold\\
  \hline
  Equal correlation	&17.06\%	&4.82\%	&0.06	&17.34\%	&0.35\%	&0.001\\
Fan \& Song's model	&6.69\%	&6.32\%	&0.0145	&6.73\%	&1.20\%	&0.001\\
Independent Cauchy	&7.12\%	&0.45\%	&0.019	&7.12\%	&0.13\%	&0.001\\
Three factor model	&5.46\%	&3.97\%	&0.014	&5.53\%	&0.31\%	&0.001\\
Two factor model	&5.00\%	&4.60\%	&0.012	&5.05\%	&0.39\%	&0.001\\
Nonlinear factor model	&6.42\%	&3.73\%	&0.019	&6.38\%	&0.68\%	&0.001\\
\hline
\end{tabular}
\end{center}
\end{table}

In Table~\ref{e5}, without loss of generality, for each dependence structure we fix threshold value 0.001 and reject the hypotheses when the dependence-adjusted $p$-values (\ref{d1}) is smaller than 0.001. Then we find the corresponding threshold value for the fixed threshold procedure such that the FDR in the two testing procedures are approximately the same. The FNR for the dependence-adjusted procedure is much smaller than that of the fixed threshold procedure, which suggests that dependence-adjusted procedure is more powerful. Note that in Table~\ref{e5}, $p_1=200$ compared with $p=1000$, implying that the better performance of the dependence-adjusted procedure is not limited to sparse situation. This is expected since subtracting common factors out make the problem have a higher signal to noise ratio.

\section{Real Data Analysis}
Our proposed multiple testing procedures are now applied to the genome-wide association studies, in particular the expression quantitative trait locus (eQTL) mapping. It is known that the expression levels of gene CCT8 are highly related to Down Syndrome phenotypes. In our analysis, we use over two million SNP genotype data and CCT8 gene expression data for 210 individuals from three different populations, testing which SNPs are associated with the variation in CCT8 expression levels. The SNP data are from the International HapMap project, which include 45 Japanese in Tokyo, Japan (JPT), 45 Han Chinese in Beijing, China (CHB), 60 Utah residents with ancestry from northern and western Europe (CEU) and 60 Yoruba in Ibadan, Nigeria (YRI). The Japanese and Chinese population are further grouped together to form the Asian population (JPTCHB). To save space, we omit the description of the data pre-processing procedures. Interested readers can find more details from the websites: http://pngu.mgh.harvard.edu/~purcell/plink/res.shtml and ftp://ftp.sanger.ac.uk/pub/genevar/, and the paper Bradic, Fan \& Wang (2010).

We further introduce two sets of dummy variables $(\mbox{\bf d}_1,\mbox{\bf d}_2)$ to recode the SNP data, where $\mbox{\bf d}_1=(d_{1,1},\cdots,d_{1,p})$ and $\mbox{\bf d}_2=(d_{2,1},\cdots,d_{2,p})$, representing three categories of polymorphisms, namely, $(d_{1,j},d_{2,j})=(0,0)$ for $\text{SNP}_j=0$ (no polymorphism), $(d_{1,j},d_{2,j})=(1,0)$ for $\text{SNP}_j=1$ (one nucleotide has polymorphism) and $(d_{1,j},d_{2,j})=(0,1)$ for $\text{SNP}_j=2$ (both nucleotides have polymorphisms). Let $\{Y^i\}_{i=1}^n$ be the independent sample random variables of $Y$, $\{d_{1,j}^i\}_{i=1}^n$ and $\{d_{2,j}^i\}_{i=1}^n$ be the sample values of $d_{1,j}$ and $d_{2,j}$ respectively. Thus, instead of using model (\ref{gwj1}), we consider two marginal linear regression models between $\{Y^i\}_{i=1}^n$ and $\{d_{1,j}^i\}_{i=1}^n$:
\begin{equation}\label{gwj2}
\min_{\alpha_{1,j},\beta_{1,j}}\frac{1}{n}\sum_{i=1}^nE(Y^i-\alpha_{1,j}-\beta_{1,j} d_{1,j}^i)^2, \ \ \ j=1,\cdots,p
\end{equation}
and between $\{Y^i\}_{i=1}^n$ and $\{d_{2,j}^i\}_{i=1}^n$:
\begin{equation}\label{gwj3}
\min_{\alpha_{2,j},\beta_{2,j}}\frac{1}{n}\sum_{i=1}^nE(Y^i-\alpha_{2,j}-\beta_{2,j} d_{2,j}^i)^2, \ \ \ j=1,\cdots,p.
\end{equation}
For ease of notation, we denote the recoded $n \times 2p$ dimensional design matrix as $\bX$. The missing SNP measurement are imputed as $0$ and the redundant SNP data are excluded. Finally, the logarithm-transform of the raw CCT8 gene expression data are used.
The details of our testing procedures are summarized as follows.
\begin{itemize}
\item
To begin with, consider the full model $Y=\alpha+\bX\beta + \epsilon$, where $Y$ is the CCT8 gene expression data, $\bX$ is the $n \times 2p$ dimensional design matrix of the SNP codings and $\epsilon_i\sim N(0,\sigma^2)$, $i=1,\cdots,n$ are the independent random errors. We adopt the refitted cross-validation (RCV) (Fan, Guo \& Hao 2010) technique to estimate $\sigma$ by $\widehat{\sigma}$, where LASSO is used in the first (variable selection) stage.
\item
Fit the marginal linear models (\ref{gwj2}) and (\ref{gwj3}) for each (recoded) SNP and obtain the least-squares estimate $\widehat{\beta}_j$ for $j=1,\cdots,2p$. Compute the values of $Z$-statistics using formula (\ref{b2}), except that $\sigma$ is replaced by $\widehat{\sigma}$.
\item
Calculate the P-values based on the $Z$-statistics and compute $R(t)=\#\{P_j: P_j \leq t\}$ for a fixed threshold $t$.
\item
Apply eigenvalue decomposition to the population covariance matrix $\bSigma$ of the $Z$-statistics. By Proposition 1, $\bSigma$ is the sample correlation matrix of $(d_{1,1},d_{2,1},\cdots,$ $d_{1,p},d_{2,p})^T$. Determine an appropriate number of factors $k$ and derive the corresponding factor loading coefficients $\{b_{ih}\}_{i=1,\ h=1}^{i=2p,\ h=k}$.
\item
Order the absolute-valued $Z$-statistics and choose the first $m=95\%\times2p$ of them. Apply $L_1$-regression to the equation set (\ref{b22}) and obtain its solution $\widehat{W}_1,\cdots,\widehat{W}_k$. Plug them into (\ref{b21}) and get the estimated $\text{FDP}(t)$.
\end{itemize}
For each intermediate step of the above procedure, the outcomes are summarized in the following figures. Figure~\ref{a61} illustrates the trend of the RCV-estimated standard deviation $\widehat{\sigma}$ with respect to different model sizes. Our result is similar to that in Fan, Guo \& Hao (2010), in that although $\widehat{\sigma}$ is influenced by the selected model size, it is relatively stable and thus provides reasonable accuracy. The empirical distributions of the $Z$-values are presented in Figure~\ref{a62}, together with the fitted normal density curves. As pointed out in Efron (2007, 2010), due to the existence of dependency among the $Z$-values, their densities are either narrowed or widened and are not $N(0,1)$ distributed. The histograms of the $P$-values are further provided in Figure~\ref{a63}, giving a crude estimate of the proportion of the false nulls for each of the three populations.
\begin{figure}
\begin{center}
\scalebox{0.44}[0.40]{\includegraphics{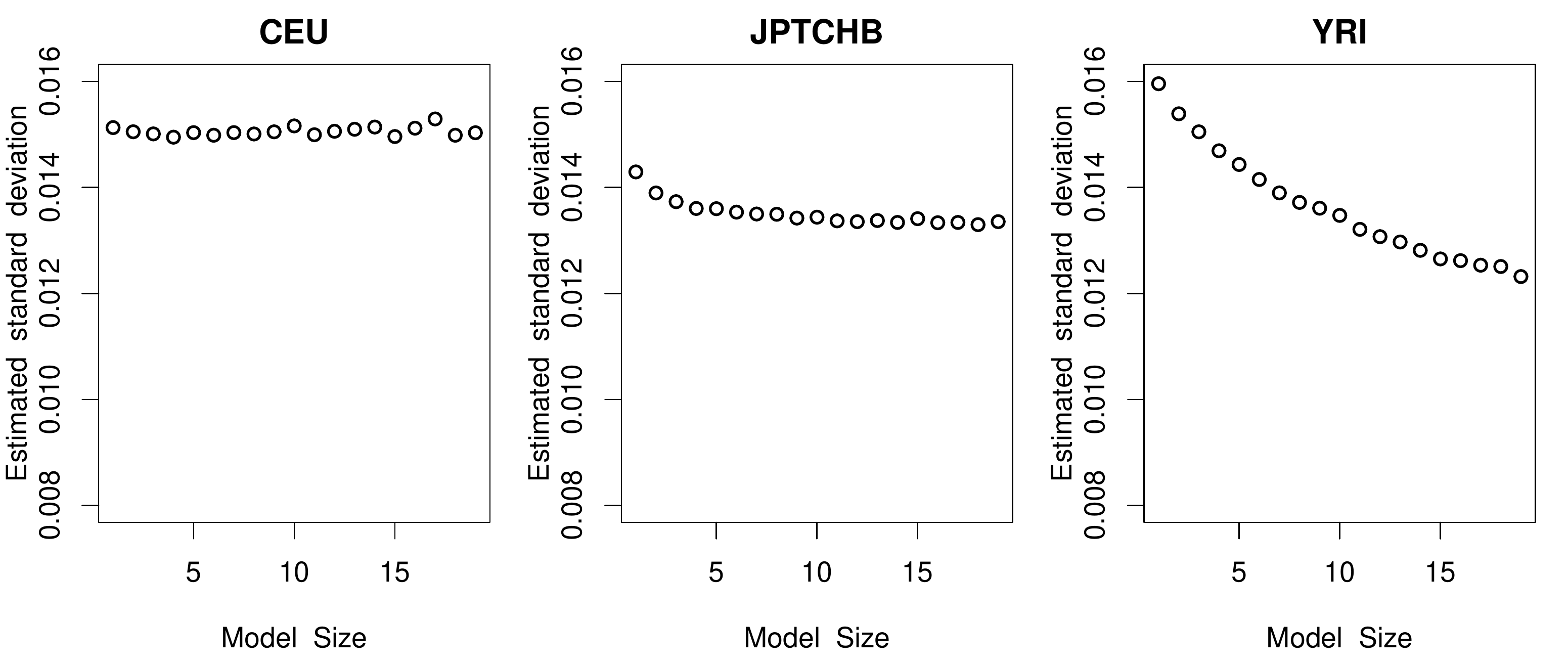}}
\end{center}
\vspace{-0.3cm}
\caption{$\widehat{\sigma}$ of the three populations with respect to the selected model sizes, derived by using refitted cross-validation (RCV).}
\label{a61}
\end{figure}
\begin{figure}[h!!!]
\begin{center}
\scalebox{0.44}[0.40]{\includegraphics{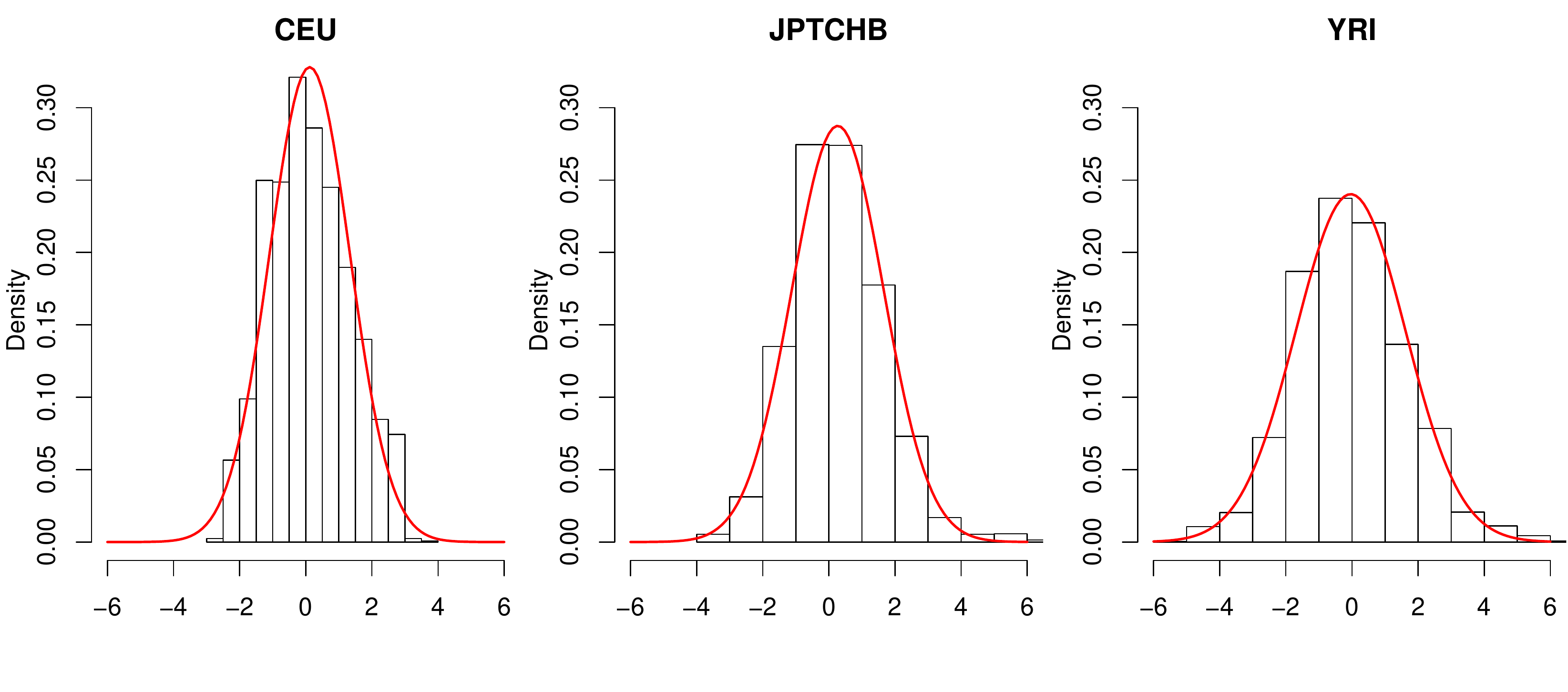}}
\end{center}
\vspace{-0.3cm}
\caption{Empirical distributions and fitted normal density curves of the $Z$-values for each of the three populations. Because of dependency, the $Z$-values are no longer $N(0,1)$ distributed. The empirical distributions, instead, are $N(0.12,1.22^2)$ for CEU, $N(0.27,1.39^2)$ for JPT and CHB, and $N(-0.04,1.66^2)$ for YRI, respectively. The density curve for CEU is closest to $N(0,1)$ and the least dispersed among the three.}
\label{a62}
\end{figure}
\begin{figure}[h!!!]
\begin{center}
\scalebox{0.44}[0.40]{\includegraphics{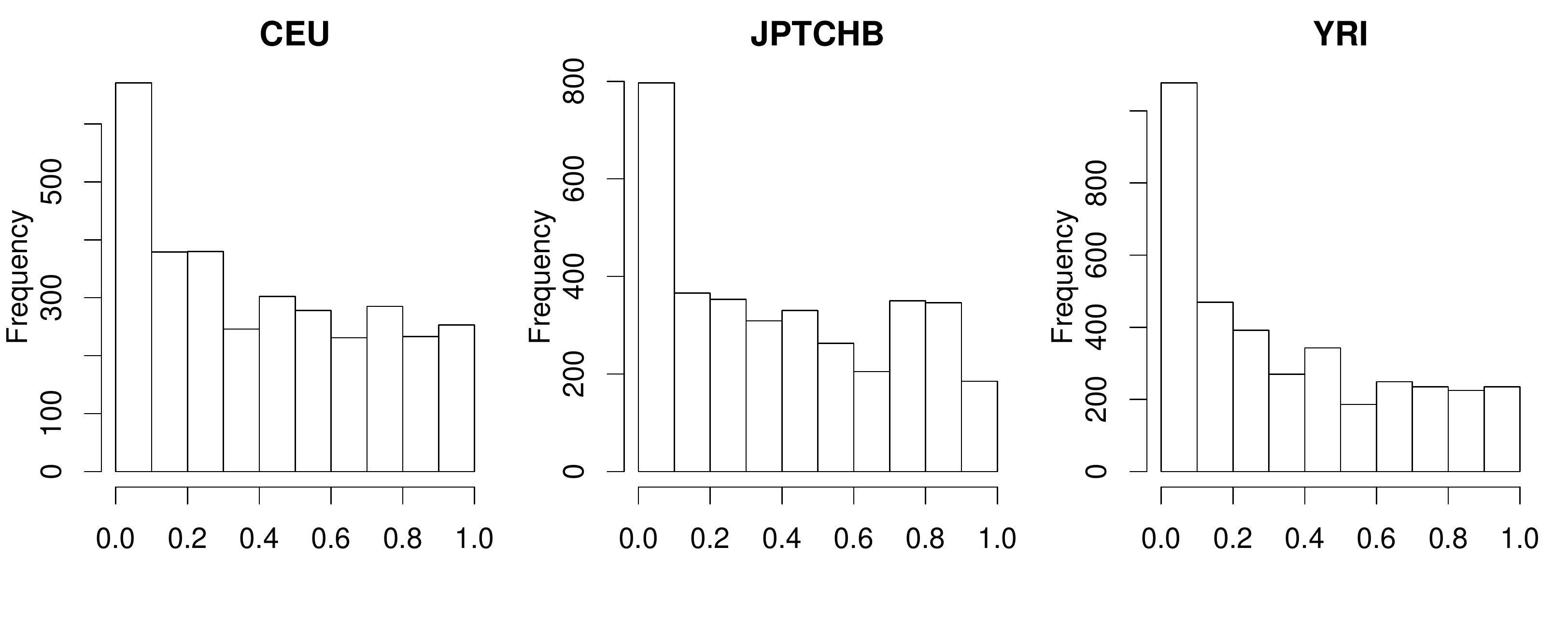}}
\end{center}
\vspace{-0.3cm}
\caption{Histograms of the $P$-values for each of the three populations. }
\label{a63}
\end{figure}
\begin{figure}[h!!!]
\begin{center}
\scalebox{0.43}[0.40]{\includegraphics{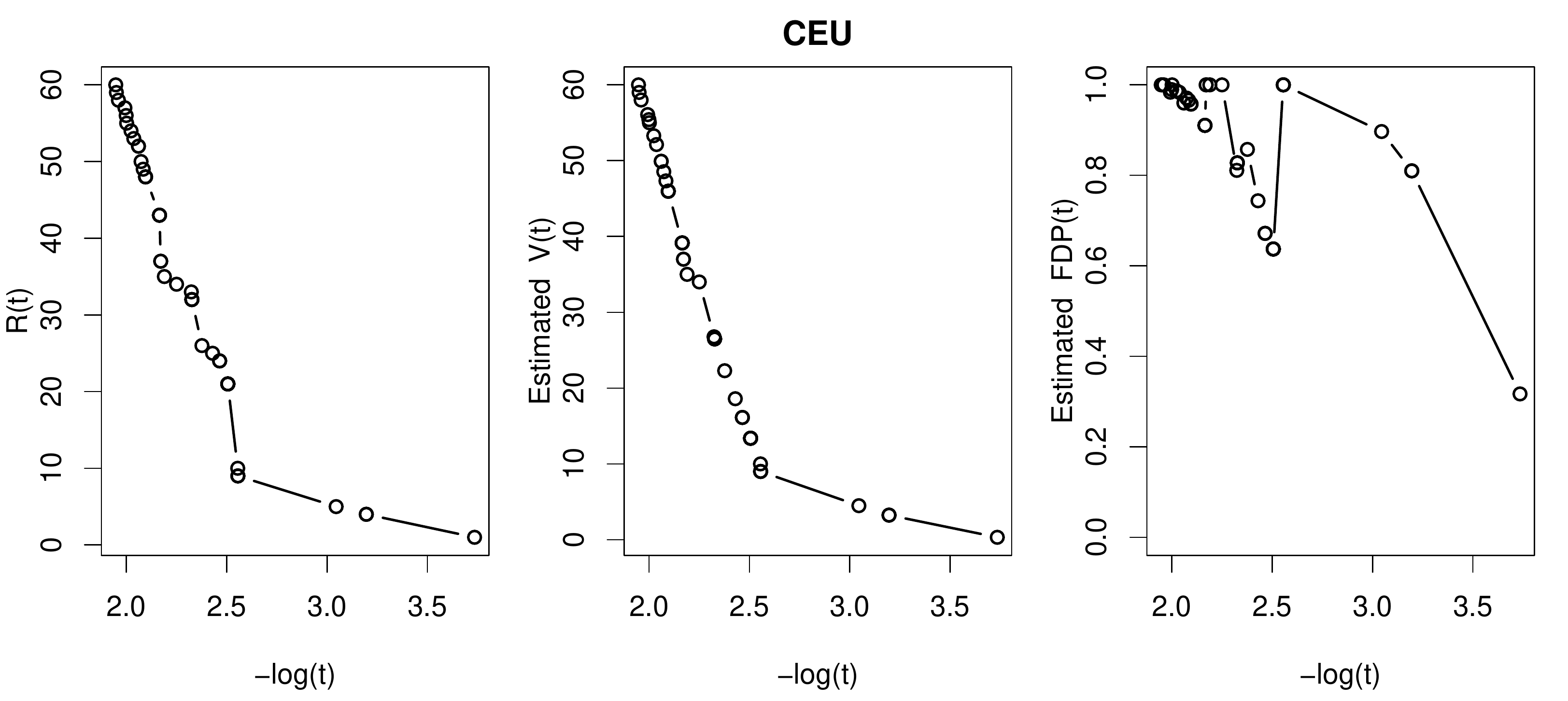}}
\scalebox{0.43}[0.40]{\includegraphics{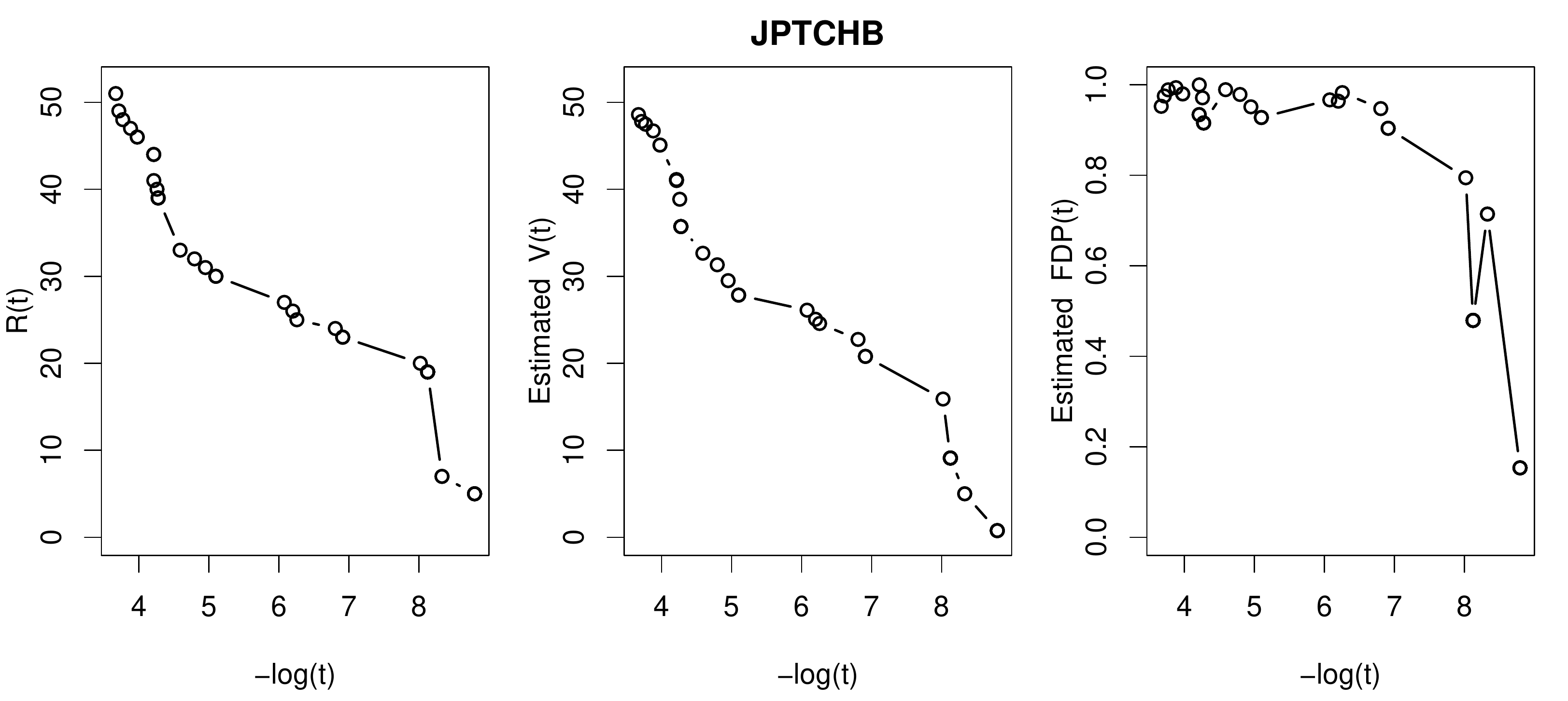}}
\scalebox{0.43}[0.40]{\includegraphics{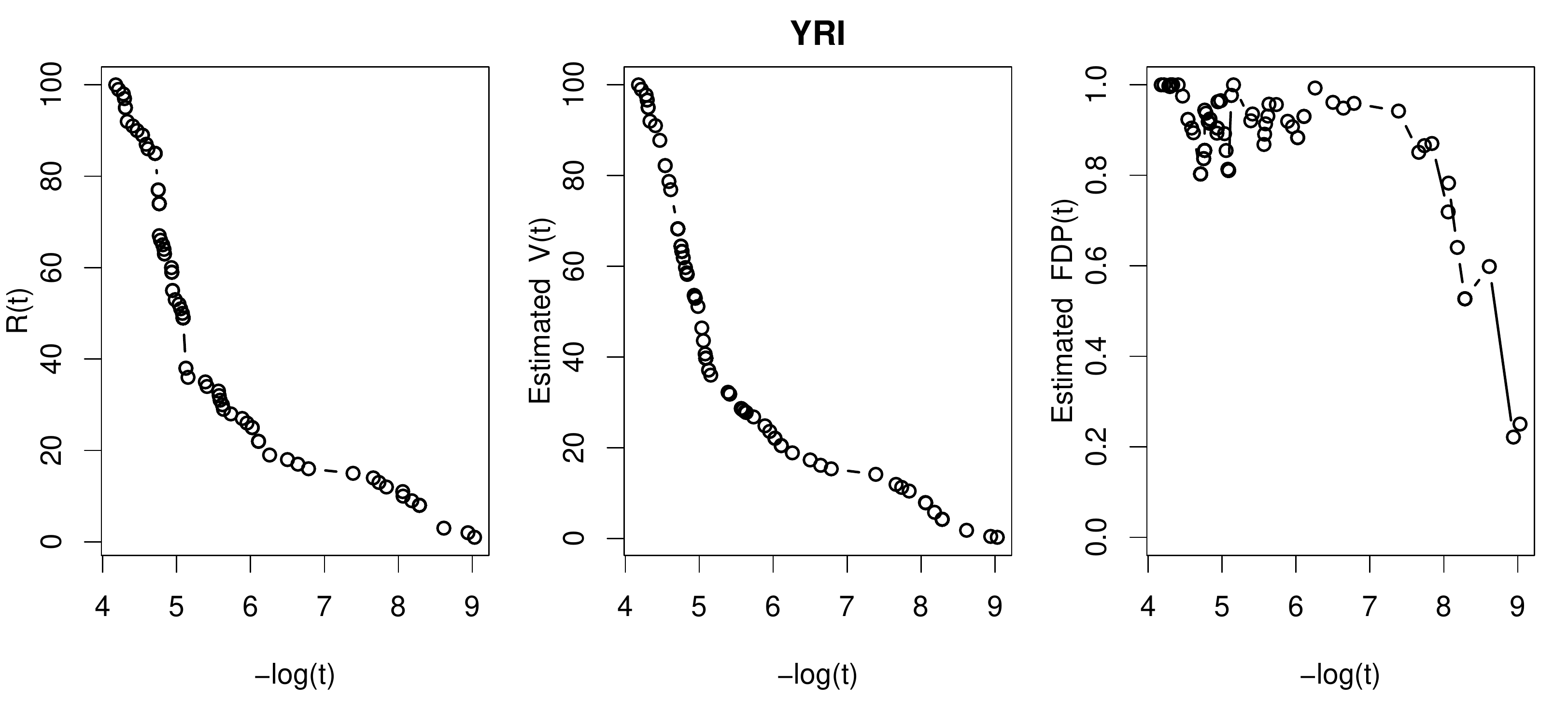}}
\end{center}
\vspace{-0.3cm}
\caption{Number of total discoveries, estimated number of false discoveries and estimated False Discovery Proportion as functions of thresholding $t$ for CEU population (row 1), JPT and CHB (row 2) and YRI (row 3). The $x$-coordinate is $-\log t$, the minus $\log_{10}$-transformed thresholding.}\label{a64}
\end{figure}
The main results of our analysis are presented in Figures~\ref{a64}, which depicts the number of total discoveries $R(t)$, the estimated number of false discoveries $\widehat{V}(t)$ and the estimated False Discovery Proportion $\widehat{\text{FDP}}(t)$ as functions of (the minus $\log_{10}$-transformed) thresholding $t$ for the three populations. As can be seen, in each case both $R(t)$ and $\widehat{V}(t)$ are decreasing when $t$ decreases, but $\widehat{\text{FDP}}(t)$ exhibits zigzag patterns and does not always decrease along with $t$, which results from the cluster effect of the P-values. A closer study of the outputs further shows that for all populations, the estimated FDP has a general trend of decreasing to the limit of around $0.1$ to $0.2$, which backs up the intuition that a large proportion of the smallest $P$-values should correspond to the false nulls (true discoveries) when Z-statistics is very large; however, in most other thresholding values, the estimated FDPs are at a high level. This is possibly due to small signal-to-noise ratios in eQTL studies.

The results of the selected SNPs, together with the estimated FDPs, are depicted in Table~\ref{a67}. It is worth mentioning that Deutsch et al. (2005) and Bradic, Fan \& Wang (2010) had also worked on the same CCT8 data to identify the significant SNPs in CEU population. Deutsch et al. (2005) performed association analysis for each SNP using ANOVA, while Bradic, Fan \& Wang (2010) proposed the penalized composite quasi-likelihood variable selection method. Their findings were different as well, for the first group identified four SNPs (exactly the same as ours) which have the smallest P-values but the second group only discovered one SNP rs965951 among those four, arguing that the other three SNPs make little additional contributions conditioning on the presence of rs965951. Our results for CEU population coincide with that of the latter group, in the sense that the false discovery rate is high in our findings and our association study is marginal rather than joint modeling among several SNPs.

\begin{table}[h!]
\begin{center}\caption{Information of the selected SNPs and the associated FDP for a particular threshold. Note that the density curve of the $Z$-values for CEU population is close to $N(0,1)$, so the approximate $\widehat{\text{FDP}}(t)$ equals $pt/R(t)\approx 0.631$. Therefore our high estimated FDP is reasonable.}\label{a67}
  \begin{tabular}{c c c c l}
  \hline\hline
   Population  & Threshold  & \# Discoveries  &  Estimated FDP  &  Selected SNPs \\
   \hline
     JPTCHB   & $1.61 \times 10^{-9}$   & 5  & $  0.1535$   &  rs965951 rs2070611\\
              &                         &    &              &  rs2832159 rs8133819\\
              &                         &    &              &  rs2832160 \\
   \hline
     YRI      & $1.14 \times 10^{-9}$   & 2  & $  0.2215$   &  rs9985076 rs965951\\
   \hline
     CEU      & $6.38 \times 10^{-4}$   & 4  & $  0.8099$   &  rs965951 rs2832159 \\
              &                         &    &              &  rs8133819 rs2832160\\
   \hline
   \end{tabular}
\end{center}
\end{table}

\begin{table}[h!]
\begin{center}\caption{Information of the selected SNPs for a particular threshold based on the dependence-adjusted procedure. The number of factors $k$ in (\ref{d1}) equals 10. The estimated FDP is based on estimator (\ref{b21}) by applying PFA to the adjusted Z-values. $*$ is the indicator for SNP equal to 2 and otherwise is the indicator for 1.}\label{a68}
  \begin{tabular}{c c c c l}
  \hline\hline
   Population  & Threshold  & \# Discoveries  &  Estimated FDP  &  Selected SNPs \\
   \hline
     JPTCHB   & $2.89 \times 10^{-4}$   & 5  & $  0.1205$   &  rs965951 rs2070611\\
              &                         &    &              &  rs2832159 rs8133819\\
              &                         &    &              &  rs2832160 \\
   \hline
     YRI      & $8.03 \times 10^{-5}$   & 4  & $  0.2080$   &  rs7283791 rs11910981\\
              &                         &    &              &  rs8128844 rs965951\\
   \hline
     CEU      & $5.16 \times 10^{-2}$   & 6  & $  0.2501$   &  rs464144* rs4817271\\
              &                         &    &              &  rs2832195  rs2831528*\\
              &                         &    &              &  rs1571671*  rs6516819*\\
   \hline
   \end{tabular}
\end{center}
\end{table}
Table~\ref{a68} lists the SNP selection based on the dependence-adjusted procedure. For JPTCHB, with slightly smaller estimated FDP, the dependence-adjusted procedure selects the same SNPs with the group selected by the fixed-threshold procedure, which suggests that these 5 SNPs are significantly associated with the variation in gene CCT8 expression levels. For YRI, rs965951 is selected by the both procedures, but the dependence-adjusted procedure selects other three SNPs which do not appear in Table~\ref{a67}. For CEU, the selections based on the two procedures are quite different. However, since the estimated FDP for CEU is much smaller in Table~\ref{a68} and the signal-to-noise ratio of the test statistics is higher from the dependence-adjusted procedure, the selection group in Table~\ref{a68} seems more reliable.

\section{Discussion}
We have proposed a new method (principal factor approximation) for high dimensional multiple testing where the test statistics have an arbitrary dependence structure. For multivariate normal test statistics with a known covariance matrix, we can express the test statistics as an approximate factor model with weakly dependent random errors, by applying spectral decomposition to the covariance matrix. We then obtain an explicit expression for the false discovery proportion in large scale simultaneous tests. This result has important applications in controlling FDP and FDR. We also provide a procedure to estimate the realized FDP, which, in our simulation studies, correctly tracks the trend of FDP with smaller amount of noise.

To take into account of the dependence structure in the test statistics, we propose a dependence-adjusted procedure with different threshold values for magnitude of $Z_i$ in different hypotheses. This procedure has been shown in simulation studies to be more powerful than the fixed threshold procedure. An interesting research question is how to take advantage of the dependence structure such that the testing procedure is more powerful or even optimal under arbitrary dependence structures.

While our procedure is based on a known correlation matrix, we would expect that it can be adapted to the case with estimated covariance matrix.  The question is then how accuracy the covariance matrix should be so that a simple substitution procedure will give an accurate estimate of FDP.

We provide a simple method to estimate the realized principal factors.  A more accurate method is probably the use of penalized least-squares method to explore the sparsity and to estimate the realized principle factor.

\section{Appendix}
Lemma 1 is fundamental to our proof of Theorem 1 and Proposition 2. The result is known in probability, but has the formal statement and proof in Lyons (1988).
\begin{lemma}[Strong Law of Large Numbers for Weakly Correlated Variables]
Let $\{X_n\}_{n=1}^{\infty}$ be a sequence of real-valued random variables such that $E|X_n|^2\leq1$. If $|X_n|\leq1$ a.s. and $\sum_{N\geq1}\frac{1}{N}E|\frac{1}{N}\sum_{n\leq N}X_n|^2<\infty$, then $\lim_{N\rightarrow\infty}\frac{1}{N}\sum_{n\leq N}X_n=0 \ \ a.s.$.
\end{lemma}
\noindent\textbf{Proof of Proposition 2:} Note that $P_i = 2\Phi(-|Z_i|)$.
Based on the expression of $(Z_1,\cdots,Z_p)^T$ in (\ref{b20}), $\Big\{I(P_i\leq t|W_1,\cdots,W_k)\Big\}_{i=1}^p$ are dependent random variables. Nevertheless, we want to prove
\begin{equation} \label{b27}
p^{-1}\sum_{i=1}^p[I(P_i\leq t|W_1,\cdots,W_k)-P(P_i\leq t|W_1,\cdots,W_k)]\stackrel{p\rightarrow\infty}{\longrightarrow}0 \ a.s. .
\end{equation}
Letting $X_i=I(P_i\leq t|W_1,\cdots,W_k)-P(P_i\leq t|W_1,\cdots,W_k)$, by Lemma 1 the conclusion (\ref{b27}) is correct if we can show
\begin{equation*}
\Var\Big(p^{-1}\sum_{i=1}^{p}I(P_i\leq t|W_1,\cdots,W_k)\Big)=O_p(p^{-\delta}) \ \ \text{for some} \ \delta>0.
\end{equation*}
\vspace{-0.1cm}
To begin with, note that
\begin{eqnarray*}
&&\Var\Big(p^{-1}\sum_{i=1}^{p}I(P_i\leq t|W_1,\cdots,W_k)\Big)\\
&=&p^{-2}\sum_{i=1}^{p}\Var\Big(I(P_i\leq t|W_1,\cdots,W_k)\Big)\\
&&+2p^{-2}\sum_{1\leq i<j\leq p}\Cov\Big(I(P_i\leq t|W_1,\cdots,W_k),I(P_j\leq t|W_1,\cdots,W_k)\Big).
\end{eqnarray*}
Since $\Var\big(I(P_i\leq t|W_1,\cdots,W_k)\big)\leq\frac{1}{4}$, the first term in the right-hand side of the last equation is $O_p(p^{-1})$. For the second term, the covariance is given by
\begin{eqnarray*}
&&P(P_i\leq t,P_j\leq t|W_1,\cdots,W_k)-P(P_i\leq t|W_1,\cdots,W_k)P(P_j\leq t|W_1,\cdots,W_k)\\
&=&P(|Z_i|<-\Phi^{-1}(t/2),|Z_j|<-\Phi^{-1}(t/2)|W_1,\cdots,W_k)\\
&&-P(|Z_i|<-\Phi^{-1}(t/2)|W_1,\cdots,W_k)P(|Z_j|<-\Phi^{-1}(t/2)|W_1,\cdots,W_k)
\end{eqnarray*}
To simplify the notation, let $\rho_{ij}^k$ be the correlation between $K_i$ and $K_j$. Without loss of generality, we assume $\rho_{ij}^k>0$ (for $\rho_{ij}^k<0$, the calculation is similar). Denote by
\begin{equation*}
c_{1,i}= a_i(-z_{t/2}-\eta_i-\mu_i), \ \ \ c_{2,i}= a_i(z_{t/2}-\eta_i-\mu_i).
\end{equation*}
Then, from the joint normality, it can be shown that
\begin{eqnarray}\label{e3}
&&P(|Z_i|<-\Phi^{-1}(t/2),|Z_j|<-\Phi^{-1}(t/2)|W_1,\cdots,W_k)\nonumber\\
&=&P(c_{2,i}/a_i<K_i<c_{1,i}/a_i, c_{2,j}/a_j<K_j<c_{1,j}/a_j)\nonumber\\
&=&\int_{-\infty}^{\infty}\Big[\Phi\Big(\frac{(\rho_{ij}^k)^{1/2}z+c_{1,i}}{(1-\rho_{ij}^k)^{1/2}}\Big)-\Phi\Big(\frac{(\rho_{ij}^k)^{1/2}z+c_{2,i}}{(1-\rho_{ij}^k)^{1/2}}\Big)\Big]\\
&&\quad\quad\times\Big[\Phi\Big(\frac{(\rho_{ij}^k)^{1/2}z+c_{1,j}}{(1-\rho_{ij}^k)^{1/2}}\Big)-\Phi\Big(\frac{(\rho_{ij}^k)^{1/2}z+c_{2,j}}{(1-\rho_{ij}^k)^{1/2}}\Big)\Big]\phi(z)dz.\nonumber
\end{eqnarray}

Next we will use Taylor expansion to analyze the joint probability further. We have shown that $(K_1,\cdots,K_p)^T\sim N(0,\bA)$ are weakly dependent random variables. Let $cov_{ij}^k$ denote the covariance of $K_i$ and $K_j$, which is the $(i,j)$th element of the covariance matrix $\bA$. We also let $b_{ij}^k=(1-\sum_{h=1}^kb_{ih}^2)^{1/2}(1-\sum_{h=1}^kb_{jh}^2)^{1/2}$. By the H\"{o}lder inequality,
\begin{eqnarray*}
p^{-2}\sum_{i,j=1}^p|cov_{ij}^k|^{1/2}\leq p^{-1/2}(\sum_{i,j=1}^p|cov_{ij}^k|^2)^{1/4}=\Big[p^{-2}(\sum_{i=k+1}^{p}\lambda_i^2)^{1/2}\Big]^{1/4}\rightarrow0
\end{eqnarray*}
as $p\rightarrow\infty$. For each $\Phi(\cdot)$, we apply Taylor expansion with respect to $(cov_{ij}^k)^{1/2}$,
\begin{eqnarray*}
\Phi\Big(\frac{(\rho_{ij}^k)^{1/2}z+c_{1,i}}{(1-\rho_{ij}^k)^{1/2}}\Big)&=&\Phi\Big(\frac{(cov_{ij}^k)^{1/2}z+(b_{ij}^k)^{1/2}c_{1,i}}{(b_{ij}^k-cov_{ij}^k)^{1/2}}\Big)\\
      &=&\Phi(c_{1,i})+\phi(c_{1,i})(b_{ij}^k)^{-1/2}z(cov_{ij}^k)^{1/2}\\
      &&\quad\quad\quad+\frac{1}{2}\phi(c_{1,i})c_{1,i}(b_{ij}^k)^{-1}(1-z^2)cov_{ij}^k+R(cov_{ij}^k).
\end{eqnarray*}
where $R(cov_{ij}^k)$ is the Lagrange residual term in the Taylor's expansion, and $R(cov_{ij}^k)=f(z)O(|cov_{ij}^k|^{3/2})$ in which $f(z)$ is a polynomial function of $z$ with the highest order as 6.

Therefore, we have (\ref{e3}) equals
\begin{eqnarray*}
&&\Big[\Phi(c_{1,i})-\Phi(c_{2,i})\Big]\Big[\Phi(c_{1,j})-\Phi(c_{2,j})\Big]\\
&&\quad\quad+\Big(\phi(c_{1,i})-\phi(c_{2,i})\Big)\Big(\phi(c_{1,j})-\phi(c_{2,j})\Big)(b_{ij}^k)^{-1}cov_{ij}^k+O(|cov_{ij}^k|^{3/2}),
\end{eqnarray*}
where we have used the fact that $\int_{-\infty}^{\infty} z\phi(z)dz=0$, $\int_{-\infty}^{\infty} (1-z^2)\phi(z)dz=0$ and the finite moments of standard normal distribution are finite. Now since $P(|Z_i|<-\Phi^{-1}(t/2)|W_1,\cdots,W_k)=\Phi(c_{1,i})-\Phi(c_{2,i})$, we have
\begin{eqnarray*}
&&\Cov\Big(I(P_i\leq t|W_1,\cdots,W_k),I(P_j\leq t|W_1,\cdots,W_k)\Big)\\
&=&\Big(\phi(c_{1,i})-\phi(c_{2,i})\Big)\Big(\phi(c_{1,j})-\phi(c_{2,j})\Big)a_ia_jcov_{ij}^k+O(|cov_{ij}^k|^{3/2}).
\end{eqnarray*}
In the last line, $\big(\phi(c_{1,i})-\phi(c_{2,i})\big)\big(\phi(c_{1,j})-\phi(c_{2,j})\big)a_ia_j$ is bounded by some constant except on a countable collection of measure zero sets. Let $C_i$ be defined as the set $\{z_{t/2}+\eta_i+\mu_i=0\}\cup\{z_{t/2}-\eta_i-\mu_i=0\}$. On the set $C_i^c$, $\big(\phi(c_{1,i})-\phi(c_{2,i})\big)a_i$ converges to zero as $a_i\rightarrow\infty$. Therefore, $\big(\phi(c_{1,i})-\phi(c_{2,i})\big)\big(\phi(c_{1,j})-\phi(c_{2,j})\big)a_ia_j$ is bounded by some constant on $(\bigcup_{i=1}^pC_i)^c$.

By the Cauchy-Schwartz inequality and $(C0)$ in Theorem 1, $p^{-2}\sum_{i,j}|cov_{i,j}^k|=O(p^{-\delta})$. Also we have $|cov_{ij}^k|^{3/2}<|cov_{ij}^k|$. On the set $(\bigcup_{i=1}^pC_i)^c$, we conclude that
\begin{equation*}
\Var\Big(p^{-1}\sum_{i=1}^pI(P_i\leq t|W_1,\cdots,W_k)\Big)=O_p(p^{-\delta}).
\end{equation*}
Hence by Lemma 1, for fixed $(w_1,\cdots,w_k)^T$,
\begin{equation}\label{g1}
p^{-1}\sum_{i=1}^p\big\{I(P_i\leq t|W_1=w_1,\cdots,W_k=w_k)
-P(P_i\leq t|W_1=w_1,\cdots,W_k=w_k)\big\}\stackrel{p\to\infty}{\longrightarrow}0\ \text{a.s.}.
\end{equation}
If we define the probability space on which $(W_1,\cdots,W_k)$ and $(K_1,\cdots,K_p)$ are
constructed as in (10) to be $(\Omega, \mathcal{F},\nu)$, with $\mathcal{F}$ and $\nu$ the
associated $\sigma-$algebra and (Lebesgue) measure, then in a more formal way, $(\ref{g1})$ is
equivalent to
\begin{equation*}
p^{-1}\sum_{i=1}^p\big\{I(P_i(\omega)\leq t|W_1=w_1,\cdots,W_k=w_k)
-P(P_i\leq t|W_1=w_1,\cdots,W_k=w_k)\big\}\stackrel{p\to\infty}{\longrightarrow}0
\end{equation*}
for each fixed $(w_1,\cdots,w_k)^T$ and almost every $\omega\in\Omega$,
leading further to
\begin{equation*}
p^{-1}\sum_{i=1}^p\big\{I(P_i(\omega)\leq t)
-P(P_i\leq t|W_1(\omega),\cdots,W_k(\omega))\big\}\stackrel{p\to\infty}{\longrightarrow}0
\end{equation*}
for almost every $\omega\in\Omega$, which is the definition for
\begin{equation*}
p^{-1}\sum_{i=1}^p\big\{I(P_i\leq t)
-P(P_i\leq t|W_1,\cdots,W_k)\big\}\stackrel{p\to\infty}{\longrightarrow}0\ \text{a.s.}.
\end{equation*}
Therefore,
\begin{equation*}
\lim_{p\to\infty}p^{-1}\sum_{i=1}^p\Big\{I(P_i\leq t)
-\big [\Phi(a_i(z_{t/2}+\eta_i+\mu_i))+\Phi(a_i(z_{t/2}-\eta_i-\mu_i))\big ]\Big\}=0\ \text{a.s.}.
\end{equation*}
With the same argument we can show
\vspace{-0.1cm}
\begin{equation*}
\lim_{p\to\infty}p_0^{-1}\Big\{V(t)-\sum_{i\in\{\text{true null}\}}
\big [\Phi(a_i(z_{t/2}+\eta_i))+\Phi(a_i(z_{t/2}-\eta_i))\big ]\Big\}=0\ \text{a.s.}
\end{equation*}
for the high dimensional sparse case. The proof of Proposition 2 is now complete.\\

\noindent\textbf{Proof of Theorem 1:}\\
For ease of notation, denote
$\sum_{i=1}^p\big [\Phi(a_i(z_{t/2}+\eta_i+\mu_i))+\Phi(a_i(z_{t/2}-\eta_i-\mu_i))\big ]$ as $\tilde R(t)$
and $\sum_{i\in\{\text{true null}\}}\big [\Phi(a_i(z_{t/2}+\eta_i))+\Phi(a_i(z_{t/2}-\eta_i))\big ]$ as
$\tilde V(t)$, then

\begin{equation*}
\begin{array}{rl}
  &\displaystyle\lim_{p\to\infty}\Big \{\text{FDP}(t)
  -\displaystyle\frac{\sum_{i\in\{\text{true null}\}}\big [\Phi(a_i(z_{t/2}+\eta_i))+\Phi(a_i(z_{t/2}-\eta_i))\big ]}
  {\sum_{i=1}^p\big [\Phi(a_i(z_{t/2}+\eta_i+\mu_i))+\Phi(a_i(z_{t/2}-\eta_i-\mu_i))\big ]}\Big\} \\
= &\displaystyle\lim_{p\to\infty}\Big\{\displaystyle\frac{V(t)}{R(t)}-\displaystyle\frac{\tilde V(t)}{\tilde R(t)}\Big\}  \\
= &\displaystyle\lim_{p\to\infty}\displaystyle\frac{(V(t)/p_0)[(\tilde R(t)-R(t))/p]+(R(t)/p)[(V(t)-\tilde V(t))/p_0]}
  {R(t)\tilde R(t)/(p_0p)}\\
= &0\ \text{a.s.}
\end{array}
\end{equation*}
by the results in Proposition 2 and the fact that both $p_0^{-1}V(t)$ and $p^{-1}R(t)$ are bounded
random variables. The proof of Theorem 1 is complete.

\noindent\textbf{Proof of Theorem 2:} Letting
\begin{eqnarray*}
\Delta_1&=&\sum_{i=1}^p\Big[\Phi(a_i(z_{t/2}+\bb_i^T\hw))-\Phi(a_i(z_{t/2}+\bb_i^T\bw))\Big]\quad\quad\text{and}\\
\Delta_2&=&\sum_{i=1}^p\Big[\Phi(a_i(z_{t/2}-\bb_i^T\hw))-\Phi(a_i(z_{t/2}-\bb_i^T\bw))\Big],
\end{eqnarray*}
we have
\begin{equation*}
\widehat{\FDP}(t)-\FDP_A(t)=(\Delta_1+\Delta_2)/R(t).
\end{equation*}
Consider $\Delta_1=\sum_{i=1}^p\Delta_{1i}$. By the mean value theorem, there exists $\xi_i$ in the interval of $(\bb_i^T\hw,\bb_i^T\bw)$, such that
$\Delta_{1i}=\phi(a_i(z_{t/2}+\xi_i))a_i\bb_i^T(\hw-\bw)$ where $\phi(\cdot)$ is the standard normal density function.

Next we will show that $\phi(a_i(z_{t/2}+\xi_i))a_i$ is bounded by a constant. Without loss of generality, we discuss about the case in (C2) when $z_{t/2}+\bb_i^T\bw<-\tau$. By (C3), we can choose sufficiently large $p$ such that $z_{t/2}+\xi_i<-\tau/2$. For the function $g(a)=\exp(-a^2x^2/8)a$, $g(a)$ is maximized when $a=2/x$. Therefore,
\begin{equation*}
\sqrt{2\pi}\phi(a_i(z_{t/2}+\xi_i))a_i<a_i\exp(-a_i^2\tau^2/8)\leq2\exp(-1/2)/\tau.
\end{equation*}
For $z_{t/2}+\bb_i^T\bw>\tau$ we have the same result. In both cases, we can use a constant $D$ such that $\phi(a_i(z_{t/2}+\xi_i))a_i\leq D$.

By the Cauchy-Schwartz inequality, we have $\sum_{i=1}^p|b_{ih}|\leq(p\sum_{i=1}^pb_{ih}^2)^{1/2}=(p\lambda_h)^{1/2}$. Therefore, by the Cauchy-Schwartz inequality and the fact that $\sum_{h=1}^k\lambda_h<p$, we have
\begin{eqnarray*}
|\Delta_{1}|&\leq&D\sum_{i=1}^p\Big[\sum_{h=1}^k|b_{ih}||\widehat{w}_h-w_h|\Big]\\
       &\leq&D\sum_{h=1}^k(p\lambda_h)^{1/2}|\widehat{w}_h-w_h|\\
       &\leq&D\sqrt{p}\Big(\sum_{h=1}^k\lambda_h\sum_{h=1}^k(\widehat{w}_h-w_h)^2\Big)^{1/2}\\
       &<&Dp\|\hw-\bw\|_2.
\end{eqnarray*}
By (C1) in Theorem 2, $R(t)/p>H$ for $H>0$ when $p\rightarrow\infty$. Therefore, $|\Delta_{1}/R(t)|=O(\|\widehat{\bw}-\bw\|_2)$. For $\Delta_{2}$, the result is the same. The proof of Theorem 2 is now complete.

\noindent\textbf{Proof of Theorem 3:} Without loss of generality, we assume that the true value of $\bw$ is zero, and we need to prove $\|\hw\|_2=O_p(\sqrt{\frac{k}{m}})$. Let $L: R^k\rightarrow R^k$ be defined by
\begin{equation*}
L_j(\bw)=m^{-1}\sum_{i=1}^m b_{ij}sgn(K_i-\bb_i^T\bw)
\end{equation*}
where $sgn(x)$ is the sign function of $x$ and equals zero when $x=0$. Then we want to prove that there is a root $\hw$ of the equation $L(\bw)=0$ satisfying $\|\hw\|_2^2=O_p(k/m)$. By classical convexity argument, it suffices to show that with high probability, $\bw^TL(\bw)<0$ with $\|\bw\|_2^2=Bk/m$ for a sufficiently large constant $B$.

Let $V=\bw^TL(\bw)=m^{-1}\sum_{i=1}^mV_i$, where $V_i=(\bb_i^T\bw)sgn(K_i-\bb_i^T\bw)$. By Chebyshev's inequality, $P(V<E(V)+h\times \SD(V))>1-h^{-2}$. Therefore, to prove the result in Theorem 3, we want to derive the upper bounds for $E(V)$ and $\SD(V)$ and show that $\forall h>0$, $\exists B$ and $M$ s.t. $\forall m>M$, $P(V<0)>1-h^{-2}$.

We will first present a result from Polya (1945), which will be very useful for our proof. For $x>0$,
\begin{equation}\label{a70}
\Phi(x)=\frac{1}{2}\Big[1+\sqrt{1-\exp(-\frac{2}{\pi}x^2)}\Big](1+\delta(x)) \ \ \ \text{with} \ \ \sup_{x>0}|\delta(x)|<0.004.
\end{equation}
The variance of $V$ is shown as follows:
\begin{equation*}
\Var(V)=m^{-2}\sum_{i=1}^m\Var(V_i)+m^{-2}\sum_{i\neq j}\Cov(V_i,V_j).
\end{equation*}
Write $\bw=s\bu$ with $\|\bu\|_2=1$ where $s=(Bk/m)^{1/2}$.  By (C5), (C6) and (C7) in Theorem 3, for sufficiently large $m$,
\begin{eqnarray}\label{d1}
\sum_{i=1}^m\Var(V_i)&=&\sum_{i=1}^mI(|\bb_i^T\bu|\leq d)\Var(V_i)+\sum_{i=1}^mI(|\bb_i^T\bu|>d)\Var(V_i)\nonumber \\
                 &=&\Big[\sum_{i=1}^mI(|\bb_i^T\bu|>d)\Var(V_i)\Big](1+o(1)),
\end{eqnarray}
and
\vspace{-0.3cm}
\begin{eqnarray}\label{d2}
\sum_{i\neq j}\Cov(V_i,V_j)&=&\sum_{i\neq j}I(|\bb_i^T\bu|\leq d)I(|\bb_j^T\bu|\leq d)\Cov(V_i,V_j)\nonumber \\
     &&+2\sum_{i\neq j}I(|\bb_i^T\bu|\leq d)I(|\bb_j^T\bu|>d)\Cov(V_i,V_j)\nonumber\\
     &&+\sum_{i\neq j}I(|\bb_i^T\bu|>d)I(|\bb_j^T\bu|>d)\Cov(V_i,V_j)\nonumber\\
     &=&\Big[\sum_{i\neq j}I(|\bb_i^T\bu|>d)I(|\bb_j^T\bu|>d)\Cov(V_i,V_j)\Big](1+o(1)).
\end{eqnarray}
We will prove (\ref{d1}) and (\ref{d2}) in detail at the end of proof for Theorem 3.

For each pair of $V_i$ and $V_j$, it is easy to show that
\begin{equation*}
\Cov(V_i,V_j)=4(\bb_i^T\bw)(\bb_j^T\bw)\Big[P(K_i<\bb_i^T\bw,K_j<\bb_j^T\bw)-\Phi(a_i\bb_i^T\bw)\Phi(a_i\bb_j^T\bw)\Big].
\end{equation*}
The above formula includes the $\Var(V_i)$ as a specific case.

By Polya's approximation (\ref{a70}),
\begin{equation}\label{d3}
\Var(V_i)=(\bb_i^T\bw)^2\exp\Big\{-\frac{2}{\pi}(a_i\bb_i^T\bw)^2\Big\}(1+\delta_j) \ \ \text{with} \ |\delta_j|<0.004.
\end{equation}
Hence
\begin{eqnarray*}
\sum_{i=1}^mI(|\bb_i^T\bu|>d)\Var(V_i)&\leq&\sum_{i=1}^ms^2\exp\Big\{-\frac{2}{\pi}(a_ids)^2\Big\}(1+\delta_j)\\
                    &\leq&2ms^2\exp\Big\{-\frac{2}{\pi}(a_{\min}ds)^2\Big\}.
\end{eqnarray*}
To compute $\Cov(V_i,V_j)$, we have
\begin{eqnarray*}
&&P(K_i<\bb_i^T\bw,K_j<\bb_j^T\bw)\\
&=&\int_{-\infty}^{\infty}\Phi\Big(\frac{(|\rho_{ij}^k|)^{1/2}z+a_i\bb_i^T\bw}{(1-|\rho_{ij}^k|)^{1/2}}\Big)\Phi\Big(\frac{\delta_{ij}^k(|\rho_{ij}^k|)^{1/2}z+a_j\bb_j^T\bw}{(1-|\rho_{ij}^k|)^{1/2}}\Big)\phi(z)dz\\
&=&\Phi(a_i\bb_i^T\bw)\Phi(a_j\bb_j^T\bw)+\phi(a_i\bb_i^T\bw)\phi(a_j\bb_j^T\bw)a_ia_jcov_{ij}^k(1+o(1)),
\end{eqnarray*}
where $\delta_{ij}^k=1$ if $\rho_{ij}^k\geq0$ and $-1$ otherwise. Therefore,
\begin{equation}\label{d4}
\Cov(V_i,V_j)= 4(\bb_i^T\bw)(\bb_j^T\bw)\phi(a_i\bb_i^T\bw)\phi(a_j\bb_j^T\bw)a_ia_jcov_{ij}^k(1+o(1)),
\end{equation}
and
\begin{eqnarray*}
&&|\sum_{i\neq j}I(|\bb_i^T\bu|>d)I(|\bb_j^T\bu|>d)\Cov(V_i,V_j)|\\
&<&\sum_{i\neq j}s^2\exp\Big\{-(a_{\min}ds)^2\Big\}a_{\max}^2|cov_{ij}^k|(1+o(1)).
\end{eqnarray*}
Consequently, we have
\begin{equation*}
\Var(V)<\frac{2}{m}s^2\exp\Big\{-\frac{2}{\pi}(a_{\min}ds)^2\Big\}a_{\max}^2\Big[\frac{1}{m}\sum_i\sum_j|cov_{ij}^k|\Big].\\
\end{equation*}
We apply (C4) in Theorem 3 and the Cauchy-Schwartz inequality to get $\frac{1}{m}\sum_i\sum_jcov_{ij}^k\leq(\sum_{i=k+1}^p\lambda_i^2)^{1/2}$ $\leq\eta^{1/2}$, and conclude that the standard deviation of $V$ is bounded by
\begin{equation*}
\sqrt{2}sm^{-1/2}\exp\Big\{-\frac{1}{\pi}(a_{\min}ds)^2\Big\}a_{\max}(\eta)^{1/4}.
\end{equation*}
In the derivations above, we used the fact that $\bb_i^T\bu\leq\|\bb_i\|_2<1$, and the covariance matrix for $K_i$ in (21) of the paper is a submatrix for covariance matrix of $K_i$ in (10).

Next we will show that $E(V)$ is bounded from above by a negative constant. Using $x(\Phi(x)-\frac{1}{2})\geq0$, we have
\begin{eqnarray*}
-E(V)&=&\frac{2}{m}\sum_{i=1}^m\bb_i^T\bw\Big[\Phi(a_i\bb_i^T\bw)-\frac{1}{2}\Big]\\
    &\geq&\frac{2ds}{m}\sum_{i=1}^mI(|\bb_i^T\bu|>d)\Big[\Phi(a_ids)-\frac{1}{2}\Big]\\
    &=&\frac{2ds}{m}\sum_{i=1}^m\Big[\Phi(a_ids)-\frac{1}{2}\Big]-\frac{2ds}{m}\sum_{i=1}^mI(|\bb_i^T\bu|\leq d)\Big[\Phi(a_ids)-\frac{1}{2}\Big].
\end{eqnarray*}
By (C5) in Theorem 3, $\frac{1}{m}\sum_{i=1}^mI(|\bb_i^T\bu|\leq d)\rightarrow0$, so for sufficiently large $m$, we have
\begin{equation*}
-E(V)\geq \frac{ds}{m}\sum_{i=1}^m\Big[\Phi(a_ids)-\frac{1}{2}\Big].
\end{equation*}
An application of (\ref{a70}) to the right hand side of the last line leads to
\begin{eqnarray*}
-E(V)\geq \frac{ds}{m}\sum_{i=1}^m\frac{1}{2}\sqrt{1-\exp\big\{-\frac{2}{\pi}(a_{\min}ds)^2\big\}}.
\end{eqnarray*}
Note that
\begin{equation*}
1-\exp(-\frac{2}{\pi}x^2)=\frac{2}{\pi}x^2\sum_{l=0}^{\infty}\frac{1}{(l+1)!}(-\frac{2}{\pi}x^2)^l>\frac{2}{\pi}x^2\sum_{l=0}^{\infty}\frac{1}{l!}(-\frac{1}{\pi}x^2)^l=\frac{2}{\pi}x^2\exp(-\frac{1}{\pi}x^2),
\end{equation*}
so we have
\begin{equation*}
-E(V)\geq \frac{d^2}{2}s^2\sqrt{\frac{2}{\pi}}a_{\min}\exp\Big\{-\frac{1}{2\pi}(a_{\min}ds)^2\Big\}.
\end{equation*}
\indent To show that $\forall h>0$, $\exists B$ and $M$ s.t. $\forall m>M$, $P(V<0)>1-h^{-2}$, by Chebyshev's inequality and the upper bounds derived above, it is sufficient to show that
\begin{equation*}
\frac{d^2}{2}s^2\sqrt{\frac{1}{\pi}}a_{\min}\exp\Big\{-\frac{1}{2\pi}(a_{\min}ds)^2\Big\}>hsm^{-1/2}\exp\Big\{-\frac{1}{\pi}(a_{\min}ds)^2\Big\}a_{\max}\eta^{1/4}.
\end{equation*}
Recall $s=(Bk/m)^{1/2}$, after some algebra, this is equivalent to show
\begin{equation*}
d^2(Bk)^{1/2}(\pi)^{-1/2}\exp\Big\{\frac{1}{2\pi}(a_{\min}ds)^2\Big\}>2h\eta^{1/4}\frac{a_{\max}}{a_{\min}}.
\end{equation*}
By (C6), then for all $h>0$, when $B$ satisfies $d^2(Bk)^{1/2}(\pi)^{-1/2}>2h\eta^{1/4}S$, we have $P(V<0)>1-h^{-2}$. Note that $k=O(m^{\kappa})$ and $\eta=O(m^{2\kappa})$, so $k^{-1/2}\eta^{1/4}=O(1)$. To complete the proof of Theorem 3, we only need to show that (\ref{d1}) and (\ref{d2}) are correct.

To prove (\ref{d1}), by (\ref{d3}) we have
\begin{equation*}
\sum_{i=1}^mI(|\bb_i^T\bu|\leq d)\Var(V_i)\leq\sum_{i=1}^mI(|\bb_i^T\bu|\leq d)s^2d^2,
\end{equation*}
and
\begin{equation*}
\sum_{i=1}^mI(|\bb_i^T\bu|   > d)\Var(V_i)\geq\sum_{i=1}^mI(|\bb_i^T\bu|   > d)s^2d^2\exp\Big\{-\frac{2}{\pi}a_{\max}^2s^2\Big\}.
\end{equation*}
Recall $s=(Bk/m)^{1/2}$, then by (C6) and (C7), $\exp\Big\{\frac{2}{\pi}a_{\max}^2s^2\Big\}=O(1)$. Therefore, by (C5) we have
\begin{equation*}
\frac{\sum_{i=1}^mI(|\bb_i^T\bu|\leq d)\Var(V_i)}{\sum_{i=1}^mI(|\bb_i^T\bu|   > d)\Var(V_i)}\rightarrow0 \ \ \text{as} \ \ m\rightarrow\infty,
\end{equation*}
so (\ref{d1}) is correct. With the same argument and by (\ref{d4}), we can show that (\ref{d2}) is also correct. The proof of Theorem 3 is now complete.

\noindent\textbf{Proof of Theorem 4:} Note that $\|\widehat{\bW}_{\LS}-\widehat{\bW}_{\LS}^*\|_2=\|(\bX^T\bX)^{-1}\bX^T\bmu\|_2$. By the definition of $\bX$, we have $\bX^T\bX=\Lambda$, where $\Lambda=\diag(\lambda_1,\cdots,\lambda_k)$. Therefore, by the Cauchy-Schwartz inequality,
\begin{equation*}
\|\widehat{\bW}_{\LS}-\widehat{\bW}_{\LS}^*\|_2=\Big[\sum_{i=1}^k\big(\frac{\sqrt{\lambda_i}\bgamma_i^T\bmu}{\lambda_i}\big)^2\Big]^{1/2}\leq\|\bmu\|_2\Big(\sum_{i=1}^k\frac{1}{\lambda_i}\Big)^{1/2}
\end{equation*}
The proof is complete.

\bibliographystyle{ims}
\bibliography{amsis-ref}

\begin{thebibliography}{99}

\bibitem[Barras, Scaillet and Wermers(2010)]{BSW10}
        Barras, L., Scaillet, O. and Wermers, R. (2010). False Discoveries in Mutual Fund Performance: Measuring Luck in Estimated Alphas. {\em Journal of Finance}, {\bf 65}, 179-216.

\bibitem [Benjamini and Hochberg(1995)]{BenjaminiH95}
		Benjamini, Y. and Hochberg, Y. (1995). Controlling the False Discovery Rate: A Practical and Powerful Approach to
		Multiple Testing. {\em Journal of the Royal Statistical Society, Series B}, {\bf 57}, 289-300.
		
\bibitem [Benjamini and Yekutieli(2001)]{BenjaminiY01}
		Benjamini, Y. and Yekutieli, D. (2001). The Control of the False Discovery Rate in Multiple Testing Under
		Dependency. {\em The Annals of Statistics}, {\bf 29}, 1165-1188.

\bibitem [Bradic(2010)]{BFW10}
		Bradic, J., Fan, J. and Wang, W. (2010). Penalized Composite Quasi-Likelihood For Ultrahigh-Dimensional Variable Selection. {\em Journal of the Royal Statistical Society, Series B}, {\bf 73(3)}, 325-349.

\bibitem [Clarke and Hall(2009)]{ClarkeH09}
		Clarke, S. and Hall, P. (2009). Robustness of Multiple Testing Procedures Against Dependence.
		{\em The Annals of Statistics}, {\bf 37}, 332-358.

\bibitem [Delattre and Roquain(2011)]{DR11}
        Delattre, S. and Roquain, E. (2011). On the False Discovery Proportion Convergence under Gaussian Equi-Correlation. {\em Statistics and Probability Letters}, {\bf 81}, 111-115.

\bibitem [Deutsch(2005)]{Deutsch05}
		Deutsch, S., Lyle, R., Dermitzakis, E.T., Attar, H., Subrahmanyan, L., Gehrig, C., Parand, L., Gagnebin, M., Rougemont, J.,
		Jongeneel, C.V. and Antonarakis, S.E. (2005). Gene expression variation and expression quantitative trait mapping of human
		chromosome 21 genes. {\em Human Molecular Genetics}, {\bf 14}, 3741-3749.

\bibitem [Efron(2007)]{Efron07}
        Efron, B. (2007). Correlation and Large-Scale Simultaneous Significance Testing.
        {\em Journal of the American Statistical Association}, {\bf 102}, 93-103.

\bibitem [Efron(2010)]{Efron10}
        Efron, B. (2010). Correlated Z-Values and the Accuracy of Large-Scale Statistical Estimates. {\em Journal of the American Statistical Association}, {\bf 105}, 1042-1055.


\bibitem [Fan, Guo and Hao(2010)]{FGH10}
        Fan, J., Guo, S. and Hao, N. (2010). Variance Estimation Using Refitted Cross-Validation in Ultrahigh Dimensional Regression. {\em Journal of the Royal Statistical Society: Series B} to appear.

\bibitem [Fan and Li(2001)]{FL01}
   Fan, J. and Li, R. (2001).  Variable selection via nonconcave
    penalized likelihood and its oracle properties.
    {\em Journal of American Statistical Association},
    {\bf 96}, 1348-1360.

\bibitem [Fan and Song(2010)]{Fan10}
        Fan, J. and Song, R. (2010). Sure Independence Screening in Generalized Linear Models with NP-Dimensionality. {\em Annals of Statistics}, {\bf 38}, 3567-3604.


\bibitem [Ferreira and Zwinderman(2006)]{FZ06}
        Ferreira, J. and Zwinderman, A. (2006). On the Benjamini-Hochberg Method. {\em Annals of Statistics}, {\bf 34}, 1827-1849.


\bibitem [Farcomeni(2007)]{Farcomeni07}
		Farcomeni, A. (2007). Some Results on the Control of the False Discovery Rate Under Dependence.
		{\em Scandinavian Journal of Statistics}, {\bf 34}, 275-297.

\bibitem [Friguet(2009)]{Friguet09}
        Friguet, C., Kloareg, M. and Causeur, D. (2009). A Factor Model Approach to Multiple Testing Under Dependence.
        {\em Journal of the American Statistical Association}, {\bf 104}, 1406-1415.


\bibitem [Finner, Dickhaus and Roters(2007)]{FDR07}
        Finner, H., Dickhaus, T. and Roters, M. (2007). Dependency and False Discovery Rate: Asymptotics. {\em Annals of Statistis}, {\bf 35}, 1432-1455.

\bibitem [Genovese and Wasserman(2004)]{Genovese04}
        Genovese, C. and Wasserman, L. (2004). A Stochastic Process Approach to False Discovery Control. {\em Annals of Statistics}, {\bf 32}, 1035-1061.

\bibitem [Leek and Storey(2008)]{LeekS08}
		Leek, J.T. and Storey, J.D. (2008). A General Framework for Multiple Testing Dependence. {\em PNAS},
		{\bf 105}, 18718-18723.

\bibitem [Lyons(1988)]{Lyons88}
        Lyons, R. (1988). Strong Laws of Large Numbers for Weakly Correlated Random Variables. {\em The Michigan Mathematical Journal}, {\bf 35}, 353-359.

\bibitem [Meinshausen(2006)]{Meinshausen06}
        Meinshausen, N. (2006). False Discovery Control for Multiple Tests of Association under General Dependence. {\em Scandinavian Journal of Statistics}, {\bf 33(2)}, 227-237.

\bibitem [Owen(2005)]{Owen05}
		Owen, A.B. (2005). Variance of the Number of False Discoveries. {\em Journal of the Royal Statistical
		Society, Series B}, {\bf 67}, 411-426.

\bibitem [Polya(1945)]{Polya45}
        Polya, G. (1945). Remarks on computing the probability integral in one and two dimensions. {\em Proceeding of the first Berkeley symposium on mathematical statistics and probability}, 63-78.

\bibitem [Portnoy(1984)]{Portnoy84}
    Portnoy, S. (1984a). Tightness of the sequence of c.d.f. processes defined from regression fractiles. {\em Robust and Nonlinear Time Series Analysis}. Springer-Verlag, New York, 231-246.

\bibitem [Portnoy, S(1984)]{P84}
    Portnoy, S. (1984b). Asymptotic behavior of M-estimators of $p$ regression parameters when $p^2/n$ is large; I. Consistency. {\em Annals of Statistics}, {\bf 12}, 1298-1309, 1984.

\bibitem [Roquain and Villers(2010)]{Roquain}
       Roquain, E. and Villers, F. (2011). Exact Calculations For False Discovery Proportion With Application To Least Favorable Configurations. {\em Annals of Statistics}, {\bf 39}, 584-612.

\bibitem [Sarkar(2002)]{Sarkar}
       Sarkar, S. (2002). Some Results on False Discovery Rate in Stepwise Multiple Testing Procedures. {\em Annals of Statistics}, {\bf 30}, 239-257.


\bibitem [Storey(2002)]{Storey02}
		Storey, J.D. (2002). A Direct Approach to False Discovery Rates. {\em Journal of the Royal Statistical Society,
		Series B}, {\bf 64}, 479-498.

				
\bibitem [Storey (2004)]{STS04}
		Storey, J.D., Taylor, J.E. and Siegmund, D. (2004). Strong Control, Conservative Point Estimation and Simultaneous
		Conservative Consistency of False Discovery Rates: A Unified Approach. {\em Journal of the Royal Statistical
		Society, Series B}, {\bf 66}, 187-205.
		

\bibitem [Sun and Cai(2009)]{SC09}
    Sun, W. and Cai, T. (2009). Large-scale multiple testing under dependency. {\em Journal of the Royal Statistical Society, Series B}, {\bf 71}, 393-424.

\end{thebibliography}
\end{document}